\documentclass[12pt]{article}

\usepackage{cite}
\usepackage{multirow}
\usepackage{comment}
\usepackage{slashed}
\usepackage{epsfig}
\usepackage{epstopdf}
\usepackage{amssymb,amsmath}
\usepackage{hyperref}
\usepackage[dvipsnames]{xcolor}

\usepackage{setspace}

\newcommand{\bea}{\begin{eqnarray}}
\newcommand{\eea}{\end{eqnarray}}

\newcommand{\nubarnu}{\raisebox{1ex}{\hbox{\tiny(}}\overline\nu\raisebox{1ex}{\hbox{\tiny)}}\hspace{-0.5ex}}

\numberwithin{equation}{section}

\begin{document}
\begin{titlepage}
%
%
\vspace*{10mm}
\begin{center}
\baselineskip 25pt 
{\Large\bf
Displaced Vertex and Disappearing Track Signatures \\ in type-III Seesaw
}
\end{center}
\vspace{5mm}
\begin{center}
{\large
Sudip Jana\footnote{sudip.jana@mpi-hd.mpg.de}, 
Nobuchika Okada\footnote{okadan@ua.edu}, and 
Digesh Raut\footnote{draut@udel.edu}
}
\end{center}
\vspace{2mm}

\begin{center}
{\it
$^{1}$Max-Planck-Institut für Kernphysik, \\ Saupfercheckweg 1, 69117 Heidelberg, Germany \\
$^{2}$ Department of Physics and Astronomy, \\ 
University of Alabama, Tuscaloosa, Alabama 35487, USA \\
$^{3}$ Bartol Research Institute, Department of Physics and Astronomy, \\
 University of Delaware, Newark DE 19716, USA
}
\end{center}
\vspace{0.5cm}
\begin{abstract}
We investigate a prospect of probing the type-III seesaw neutrino mass generation mechanism at various collider experiments by searching for a disappearing track and a displaced vertex signature originating from the decay of $SU(2)_L$ triplet fermion ($\Sigma$).  
Since $\Sigma$ is primarily produced at colliders through the electroweak gauge interactions, its production rate is uniquely determined by its mass. 
We find that a $\Sigma$ particle with a mass of a few hundred GeV produces a disappearing track signature from the decay of its charged component, which can be searched at the HL-LHC. 
Furthermore, we show that if the lightest observed neutrino has a mass of around $10^{-9}$ eV, the neutral component of ${\Sigma}$ can be discovered at the proposed MATHUSLA detector.   
We also show that the charged component of ${\Sigma}$ with mass of a few hundred GeV can be observed at the LHeC and FCC-he as a displaced vertex signature.  
\end{abstract}
\end{titlepage}

\section{Introduction}
\label{sec:1}
The Standard Model (SM) of particle physics is a tremendously successful theory, but it is incomplete in its current form. 
Among its various shortcomings, 
we here focus on the fact that the SM offers no explanation about the origin of the tiny observed masses of the observed neutrinos \cite{numass}. 
One of the most appealing scenario to naturally generate the tiny neutrino masses is the seesaw mechanism, namely, type-I seesaw \cite{TypeI-1, TypeI-2}, type-II seesaw \cite{TypeII} and type-III seesaw \cite{TypeIII}, which effectively generates the lepton number violating dimension five operator $\mathcal{O}_5=\frac{c}{\Lambda}LLHH$ at low energies, where $\Lambda$ is the seesaw scale.

In this paper, we focus on the type-III seesaw, 
where in addition to the SM particles, 
at least two $SU(2)_L$ triplet fermions ($\Sigma$) with zero hypercharge are introduced \cite{TypeIII}. 
For example, if the type-III seesaw mechanism is incorporated in the grand unified theory (GUT) framework, the triplet fermion mass $M_\Sigma$ are of the intermediate scale \cite{TypeIII, Ma, BS, P1}\footnote{For a discussion of GUT scenario with $M_\Sigma \lesssim {\cal O} (1)$ TeV, see Ref.~\cite{collider2}.}.  
However, 
the type-III seesaw scenario with $M_\Sigma \simeq {\cal O} (1)$ TeV is also technically natural. 
Unlike GUT scenarios, 
the latter can be explored at collider experiments. See, for example, Refs.~\cite{collider, collider1, collider2, collider3}.
The CMS \cite{cms} and the ATLAS collaborations \cite{atlas} for the Large Hadron Collider (LHC) have carried out a dedicated search for an $SU(2)_L$ triplet fermion of the type-III seesaw. 
Particularly, their search focuses on the prompt decay of the neutral ($\Sigma^0$) and the charged ($\Sigma^{\pm}$) components of $\Sigma$ produced via an intermediate electroweak gauge bosons 
that yields two final-state leptons (electrons or muons) of different flavors and charge combinations  accompanied by at least two jets. 
Assuming flavor-universal branching fractions of $\Sigma^{0,\pm}$ into different lepton flavors, 
$M_\Sigma < 840$ GeV is excluded \cite{cms}.   
The direct search for promptly decaying $\Sigma^{0, \pm}$ with mass in the TeV range is very challenging at the LHC.

We explore an interesting possibility that $\Sigma^{0,\pm}$ are long-lived and evade the prompt decay searches at the LHC, thereby explaining the null results from the LHC experiments.  
The mass-splitting between $\Sigma^{\pm}$ and $\Sigma^{0}$ is small, $ {\cal O} (10^2)$ MeV, such that the pion/SM leptons produced in $\Sigma^{\pm}$ decay to $\Sigma^0$ are soft and very challenging to reconstruct at the LHC (a proton and proton ($pp$)-collider) due to a large hadronic background. 
If the neutral fermion $\Sigma^0$ lives long enough to pass through the detector undetected, the observed track of the charged $\Sigma^{\pm}$ disappear after they decay inside the detector. 
However, a detailed analysis of such a scenario has not been considered in the literature. 
We will show that this so-called disappearing track signature can be the primary discovery mode of the long-lived $\Sigma^{\pm}$ at the LHC and the future HL-LHC. 
For comparison, a dedicated disappearing track search for chargino in supersymmetric models, whose production and decay mechanism is very similar to $\Sigma^{\pm}$, has already led to a very stringent bound \cite{Aaboud:2017mpt}. 

For a long-lived charge neutral particle, its  decay vertex is located away from the collision point. 
This so-called displaced vertex signature is a very clean with almost zero SM background. 
The current status of the displaced vertex searches at the LHC can be found in Refs.~\cite{ATLAS:2012av, Chatrchyan:2012sp, Aad:2012kw, Aad:2012zx, Chatrchyan:2012jna, Aad:2014yea, CMS:2014wda, CMS:2014hka, Aaij:2014nma, Aad:2015asa, Aad:2015uaa, Aad:2015rba,Aaboud:2016dgf, Aaboud:2016uth, ATLAS:2016jza, ATLAS:2016olj, Aaboud:2017iio}. 
We expect a dramatic improvement in the search for the displaced vertex at future collider experiments, 
such as the proposed MAssive Timing Hodoscope for Ultra Stable neutraL pArticles (MATHUSLA) experiment\cite{Chou:2016lxi} of the HL-LHC\footnote{Other  dedicated displaced vertex search experiments at the LHC such as FASER \cite{Ariga:2018uku} and the proposed Codex-b (see, for example, Ref.~\cite{Dercks:2018eua}) specialize in search of long-lived neutral particle with mass of a few GeV or smaller which are created from the decay of mesons that are copiously produced at the LHC. Compared to FASER and Codex-b, MATHUSLA has the best search reach because of its large detector volume. For comparision, see, for example, Ref.~\cite{Dercks:2018eua}. }, the Large Hadron electron Collider (LHeC) \cite{AbelleiraFernandez:2012cc} and the Future Circular electron-hadron Colliders (FCC-he)\cite{Kuze:2018dqd}. 
The MATHUSLA detector is an external detector purposed to detect the decay of long-lived particles which have escaped the detection at the LHC such as $\Sigma^0$ of the type-III seesaw\footnote{Such a prospect for type-I seesaw scenario at MATHUSLA experiment has been investigated in Refs~\cite{dvn1, dvn2}.}. 
Being purposefully placed far away from the LHC, this experiment is ideal to search for a long-lived $\Sigma^0$.  
In fact, we find that MATHUSLA can discover $\Sigma^0$.  
Compared to the LHC, electron and proton ($ep$)-colliders such as the LHeC and FCC-he have cleaner environments because they are free from large hadronic background and pile-up events at higher luminosities. 
This enables explicit reconstruction of the soft final states in $\Sigma^{\pm}$ decay to $\Sigma^0$ for a displaced vertex signature.  
See, for example, Ref.~\cite{Curtin:2017bxr} for the discussion of Higgsino search prospect at the LHeC and FCC-he. 
Because of the asymmetric beam setup of the $ep$-colliders, the charged Higgsnos  produced at these colliders are boosted and effectively have a longer lifetime in the lab frame. 
Thus, $ep$-collider can access lifetimes many orders of magnitude shorter than the $pp$-colliders. 
We will show that LHeC and FCC-he can also  probe the type-III seesaw mechanism.

In the following, we will show that the disappearing track and displaced vertex signatures that arise from the decay of $\Sigma^{\pm,0}$ in type-III  seesaw can be probed at HL-LHC, MATHUSLA, LHeC and FCC-he experiments.  
This work is organized as follows: In Sec.~\ref{sec:model}, 
we introduce the type-III seesaw mechanism and discuss the neutrino mass generation mechanism. 
In Sec.~\ref{sec:production}, 
we study the production of $\Sigma^{\pm,0}$ at HL-LHC, LHeC and FCC-he, and also calculate their decay lengths. 
After discussing the bounds from the disappearing track search at the LHC experiment as well its search prospect at the HL-LHC, 
we examine in Sec.~\ref{sec:dv} the prospect of observing the displaced vertex signal of $\Sigma^{\pm,0}$ decay at  MATHUSLA, LHeC and FCC-he. 
Finally, we present our conclusion in  Sec.~\ref{sec:con}.

\section{The type-III seesaw}\label{sec:model}
We consider the extension of the SM with three generation of $S U(2)_{L}$ triplet (right-handed) fermions, $\Sigma_i$ ($i = 1, 2, 3$), and the Lagrangian is given by 
\bea
\mathcal{L}_{\Sigma}=& 
{\rm Tr}\left[{\overline \Sigma}_i {\slashed D} \Sigma_i\right]
- \left( \frac{1}{2} M_{\Sigma}^{ij} {\rm Tr}
\left[\overline{\Sigma_c^{i}}  \Sigma_j\right] + {\rm h.c}\right) 
-\left( Y_{\Sigma }^{ij} {\overline L}_i  \Sigma_j {H}+ {\rm h.c}\right), 
\label{lag}
\eea
where $D$ is the covariant derivative for $\Sigma_i$,  $M_\Sigma^{ij}$ are Majorana masses of the triplet fermions,  $Y_{\Sigma}^{ij}$ are the Dirac-type Yukawa couplings, 
$L_i \equiv(\nu_i, l_i)^{T}$ is the left-handed SM $SU(2)_L$ lepton doublet, 
$H 
\equiv
\left(H^{0}, H^{-}\right)^{T} 
=  
\left((v+h+i \eta) / \sqrt{2}, H^{-}\right)^{T}$ 
and $v= 246$ GeV. 
The $SU(2)_L$ triplet fermion contains charged ($\Sigma^{\pm}$) and charge neutral ($\Sigma^0$) components which can be expressed as
\bea
\Sigma_i=\left(\begin{array}{cc}{\Sigma^{0_i}_i / \sqrt{2}} & {\Sigma^{+}_i} \\ {\Sigma^{-}_i} & {-\Sigma^{0}_i / \sqrt{2}}\end{array}\right),
\eea
and its conjugate is given by $\Sigma^{c}_i \equiv C \bar{\Sigma_i}^{T}$, where $C$ denotes the charge conjugation operator.

\begin{figure}[t]
\begin{center}
 \includegraphics[scale=0.7]{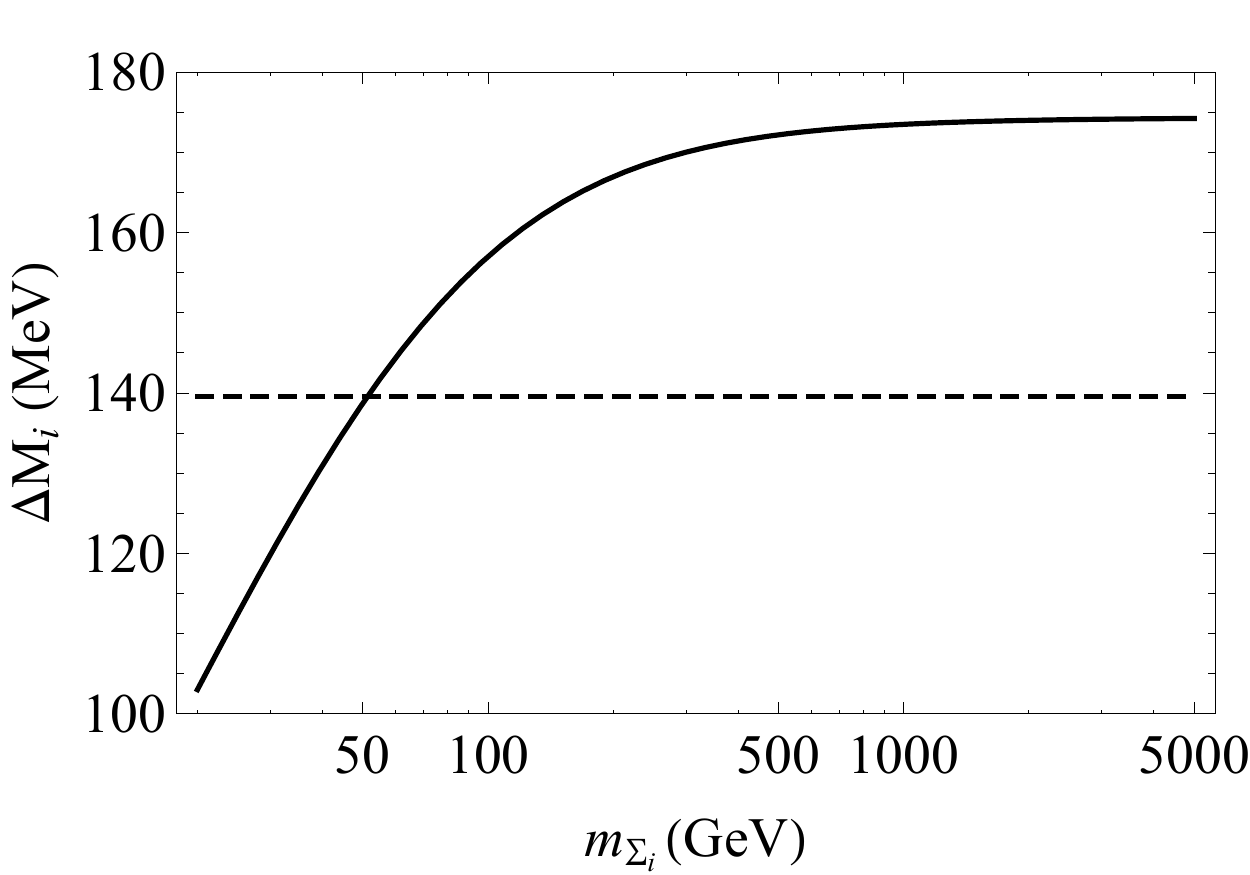}
\end{center}
 \caption{The mass-splitting between $\Sigma^{\pm}$ and $\Sigma^{0}$ induced by radiative corrections as a function of $m_{\Sigma}$. Dashed line denotes the $\pi$ meson mass.
 }
\label{fig:masssplitting}
\end{figure}

Let us first consider the masses of $\Sigma^{\pm}_i$ and $\Sigma^{0}_i$.  
Although their masses are degenerate at the tree-level, radiative corrections induced by the electroweak  gauge boson loops remove the degeneracy and generate mass-splitting between $\Sigma^{\pm}_i$ and $\Sigma^{0}_i$. 
The resulting mass difference ($\Delta M_i$) at the one-loop order is given by \cite{Cirelli:2005uq}
\begin{equation}
\Delta M_i=
\frac{\alpha_{2} m_{\Sigma_i}}{4 \pi}
\left[
f \left(\frac{M_{W}}{m_{\Sigma_i}}\right)
-\cos^2\theta_W  \times
f \left(\frac{M_{Z}}{m_{\Sigma_i}}\right) 
\right], 
\end{equation}
where $\alpha_2 = 0.034$, 
$\cos^2\theta_W \simeq 0.769$ \cite{PDG}, 
the function $f$ is defined as 
\begin{equation}
f(r)=r\left[2 r^{3} \ln r-2 r+\left(r^{2}-4\right)^{1 / 2}\left(r^{2}+2\right) \ln A\right] / 2,
\end{equation}
with $A=\left(r^{2}-2-r \sqrt{r^{2}-4}\right) / 2$.
We show $\Delta M_i$ as a function of $m_{\Sigma_i}$ in  Fig.~\ref{fig:masssplitting}. 
The dashed line denotes the $\pi$ meson mass of $140$ MeV. 

\begin{figure}[t]
  \centering
   \includegraphics[scale=0.7]{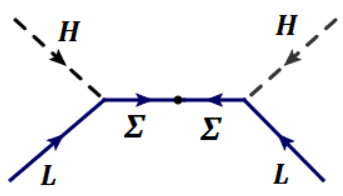}
  \caption{Tree-level neutrino mass diagram for type-III seesaw mechanism.}
      \label{fig:d5diags}
\end{figure}

The Yukawa couplings between $H$ and $\Sigma$ generate mass-mixings between $\Sigma^{0,\pm}$, $\nu_{\ell}$ and $\ell_{L}$.  
The mass matrices involving the charged fermion states, $M_\pm$ in ($l^{\pm},\Sigma^{\pm}$) basis and neutral fermion states, $M_0$ in ($\nu,\Sigma^0$) basis, are expressed as 
\bea
M^\pm =
  \left(
  \begin{array}{cc}
 m_{\ell^\pm}& -{\sqrt 2} m_D \\ 
   0 &M_{\Sigma}
  \end{array}
\right),
\qquad
  M^0 =
  \left(
  \begin{array}{cc}
   0& -m_D \\ 
  -m_D^T & M_{\Sigma}
  \end{array}
\right), 
\label{eq:mzeroIII}
\eea
respectively.  
Here, $m_{\ell^\pm}$ is the mass matrix for the SM charged leptons, $m_D^{ij} = Y_\Sigma^{ij} v/{\sqrt 2}$ are the components of Dirac type neutrino mass matrix ($m_D$) and ${M_\Sigma} =  {\rm diag} \left({m_{\Sigma_1}},{ m_{\Sigma_2}},{ m_{\Sigma_3}}\right)$ is the Majorana mass matrix for the triplet fermions. 
We have fixed the mass eigenvalues of the charged and the neutral components of each triplets to be the same because of the tiny mass-splitting between them.   
Equation~(\ref{eq:mzeroIII}) shows that the mixings between heavy triplet and light SM states in both the neutral and the charged fermion sectors are of the same order $m_DM_{\Sigma}^{-1}$. 
In the following, we will focus on the mixing in the neutral sector which generates the observed neutrino masses. 
The Feynman diagram for the type-III seesaw is shown in Fig. \ref{fig:d5diags}. 
The light neutrino mass matrix is generated as 
\bea
{m_\nu} \simeq - m_D \left(M_\Sigma\right)^{-1} m_D^T. 
\label{eq:seesaw1}
\eea
We can express the neutrino flavor eigenstate ($\nu$) as 
$\nu \simeq \mathcal{N} \nu_m+\mathcal{R} N_m$, 
where $\nu_m$ ($\Sigma_m$) are light (heavy) mass eigenstates, 
$\mathcal{R} =m_D (M_\Sigma)^{-1}$, $\mathcal{N}=\Big(1-\frac{1}{2}\mathcal{R}^\ast\mathcal{R}^T\Big)U_{\rm{MNS}}\simeq U_{\rm{MNS}}$. 
The neutrino mass mixing matrix $U_{\rm{MNS}}$ diagonalizes the light neutrino mass matrix as follows: 
\bea
  U_{\rm MNS}^T m_\nu U_{\rm MNS}  = D_\nu = {\rm diag}(m_1, m_2, m_3), 
\label{eq:seesaw2}
\eea  
where 
\bea
U_{\rm{MNS}} = \begin{pmatrix} c_{12} c_{13}&c_{12}c_{13}&s_{13}e^{-i\delta}\\-s_{12}c_{23}-c_{12}s_{23}s_{13}e^{i\delta}&c_{12}c_{23}-s_{12}s_{23}s_{13}e^{i\delta}&s_{23} c_{13}\\ s_{12}c_{23}-c_{12}c_{23}s_{13}e^{i\delta}&-c_{12}s_{23}-s_{12}c_{23}s_{13}e^{i\delta}&c_{23}c_{13}
\end{pmatrix} 
\begin{pmatrix}
1&0&0\\
0&e^{-i \rho_1}&0\\
0&0&e^{-i \rho_2}
 \end{pmatrix}. 
\eea
In this work, for simplicity, we set the Majorana $CP$-phases $\rho_{1,2} =0$  while we use the best fit values for neutrino masses, mixing angles ($c_{ij}\equiv \cos\theta_{ij}$ and $s_{ij}\equiv \sin\theta_{ij}$) and the Dirac $CP$-phase ($\delta$) from NuFIT's global analysis \cite{Esteban:2020cvm} of neutrino oscillation data.  
Their values depend on the observed neutrino mass hierarchy, namely, normal hierarchy (NH), $m_1< m_2< m_3$, and inverted hierarchy (IH), $m_3< m_1< m_2$. 
For the NH (IH), the best fit central values are $\theta_{12}\simeq33.4^{\circ}(33.5^{\circ})$, $\theta_{23}\simeq49.2^{\circ}(49.3^{\circ})$, $\theta_{13}\simeq8.57^{\circ} (8.60^{\circ})$, $\delta \simeq 197^{\circ} (282^{\circ})$, $\Delta m_{21}^2 \simeq m_2^2-m_1^2 = 7.42 (7.42) \times 10^{-5}$ eV$^2$, and 
$m_{32}^2\simeq m_3^2-m_2^2= 2.51 (-2.50) \times 10^{-3}$ eV$^2$, respectively.

From Eqs.~(\ref{eq:seesaw1}) and (\ref{eq:seesaw2}), 
the Dirac mass matrix can be parameterized as \cite{Casas:2001sr}
\bea 
 i m_D =  U_{\rm{MNS}}^* \sqrt{D_{\nu}} \; O \sqrt{M_\Sigma}, 
\label{mD}
\eea
where $\sqrt{D_\nu} = {\rm diag} \left({\sqrt m_1}, {\sqrt m_2}, {\sqrt m_3}\right)$, 
${\sqrt M_\Sigma} =  {\rm diag} \left({\sqrt m_{\Sigma_1}},{\sqrt m_{\Sigma_2}},{\sqrt m_{\Sigma_3}}\right)$ and $O$ is a $3 \times 3$ orthogonal matrix.  
A general $3\times 3$ orthogonal matrix $O$ is given by  
\bea
O =
\begin{pmatrix}
1&0&0\\
0&\cos \theta_1& \sin \theta_1 \\
0&-\sin \theta_1& \cos \theta_1 \\
\end{pmatrix}
\begin{pmatrix}
\cos \theta_2& 0 &\sin \theta_2 \\
0&1&0 \\
-\sin \theta_2&0&\cos \theta_2 \\
\end{pmatrix} 
\begin{pmatrix}
\cos \theta_3& \sin \theta_3 &0 \\
-\sin \theta_3& \cos \theta_3&0 \\
0&0&1
\end{pmatrix}, 
\eea 
where $\theta_{1,2,3}$ are generally complex valued.

\section{Production and search for fermion triplets at current and future colliders}
\label{sec:production}

\begin{figure}[t]
\begin{center}
\includegraphics[scale=0.6]{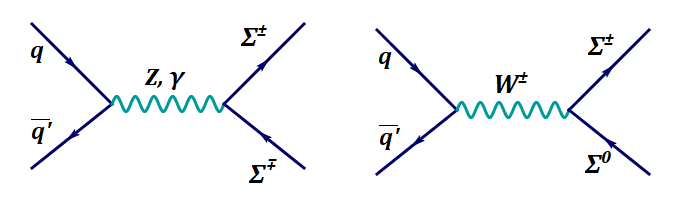} \;\;\;
\includegraphics[scale=0.6]{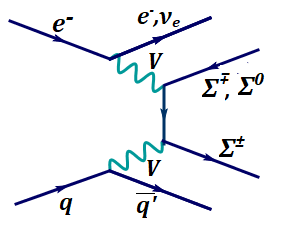}  
\end{center}
 \caption{Representative Feynman diagrams for the production of $\Sigma^{\pm}$ and $\Sigma^0$ at colliders. 
 }
\label{fig:prod}
\end{figure}

Let us discuss the production of $\Sigma^{0,\pm}$ at various collider experiments. 
Since $\Sigma$ is an $SU(2)_L$ triplet, in the type-III seesaw scenario, $\Sigma^{0,\pm}$ are produced at colliders through their electroweak gauge interactions as shown in Fig.~\ref{fig:prod}. 
All these processes only involve SM gauge interactions, and therefore $\Sigma^{0,\pm}$ production cross sections are solely determined by the triplet mass ($m_{\Sigma}$). 

The Drell-Yan processes dominate the $\Sigma$ production at a $pp$-collider such as the LHC. 
The singly charged fermion $\Sigma^{\pm}$ is pair-produced from a $q\bar{q}$ fusion through the $s$-channel $Z/\gamma$ exchange as shown in the left diagram of Fig.~\ref{fig:prod}. 
Similarly, 
the $s$-channel $W^\pm$ exchange produces $\Sigma^\pm$ and $\Sigma^0$ as shown in the middle diagram of Fig.~\ref{fig:prod}. 
The $\Sigma^{\pm}$ can also be pair-produced via vector boson-fusion process in association with two forward jets (its diagram is similar to the right diagram in Fig.~\ref{fig:prod}). 
Despite a large production rate, this process suffers from larger QCD background due to the forward jets activity and is sub-dominant\footnote{Similarly, the photon initiated vector-boson processes \cite{photon} contribution is sub-dominant.} compared to the Drell-Yan processes. 
Hence, we do not include vector boson-fusion process in our LHC analysis. 
On the other hand, the vector boson-fusion process shown in the right diagram of Fig.~\ref{fig:prod} dominates the production at the $ep$-colliders such as the LHeC and FCC-he. 
Because of the clean (low pile-up) environment of the $ep$-colliders, as we have discussed earlier, the LHeC and FCC-he can probe much shorter lifetimes and are   complementary to LHC disappearing track search of $\Sigma^\pm$ decay. 
The LHeC and FCC-he will look for the displaced vertex signature from $\Sigma^\pm$ decay.

\begin{figure}[t!]
\begin{center}
\includegraphics[width=0.49\textwidth, height=5.5cm]{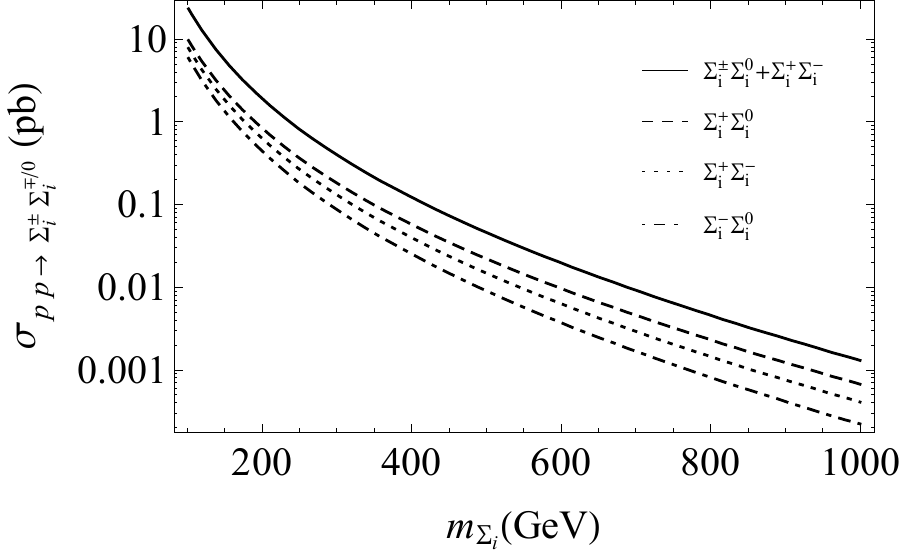}\;
\includegraphics[width=0.49\textwidth, height=5.5cm]{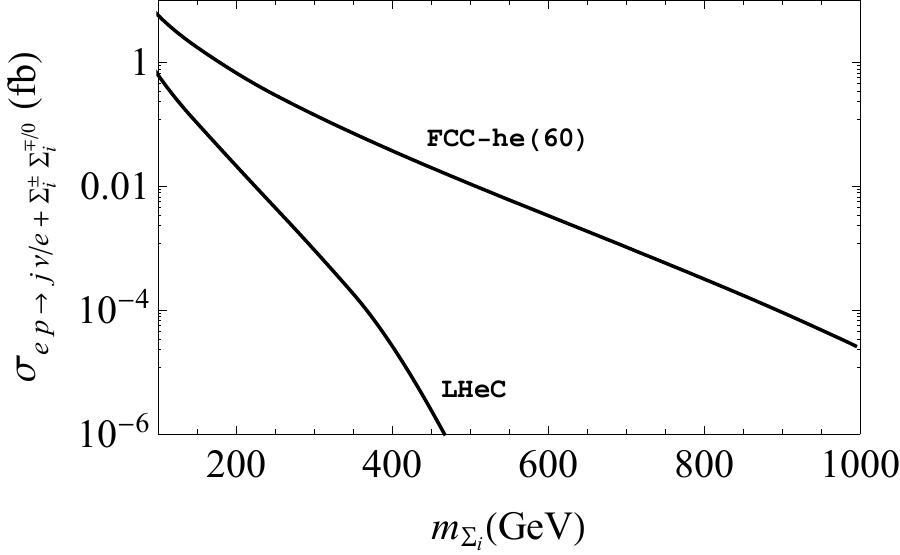}
\end{center}
 \caption{The left (right) shows the production cross section of $\Sigma^\pm_i$ and $\Sigma^0_i$ at $pp$ ($ep$)-collider as a function of its mass $m_{\Sigma_i}$.}
\label{fig:epcs}
\end{figure}

For our analysis, we generate the signal sample for fermion triplet with {\sc MadGraph5aMC@NLO}~\cite{Alwall:2011uj,Alwall:2014hca}  event-generator and evaluate the production cross section of the $\Sigma^{0,\pm}$ at LHC, LHeC and FCC-he.  
Our results are shown in Fig.~\ref{fig:epcs} as a function of $m_\Sigma$. 
In the left panel of Fig.~\ref{fig:epcs}, 
we show the production cross section for the processes at the LHC with the center of mass energy ${\sqrt s} = 13$ TeV. 
In the right panel of Fig.~\ref{fig:epcs}, 
we show the production cross section for the 
LHeC (FCC-he) collider, respectively, 
where the proton beam energy is set to be 7 (50) TeV while the $e^-$ beam energy is set to be $60$ GeV for both cases.

\begin{figure}[t!]
\begin{center}
\includegraphics[width=0.7\textwidth, height=3cm]{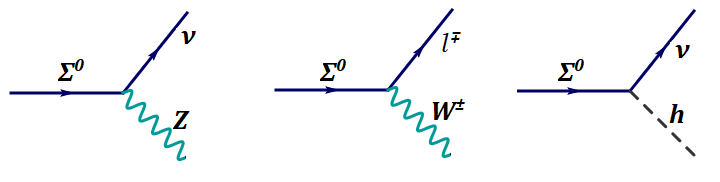} 
\end{center}
 \caption{Representative Feynman diagrams for $\Sigma^{0}$ decay. 
 }
\label{fig:decay2}
\end{figure}

\begin{figure}[t!]
\begin{center}
\includegraphics[width=0.7\textwidth, height=5cm]{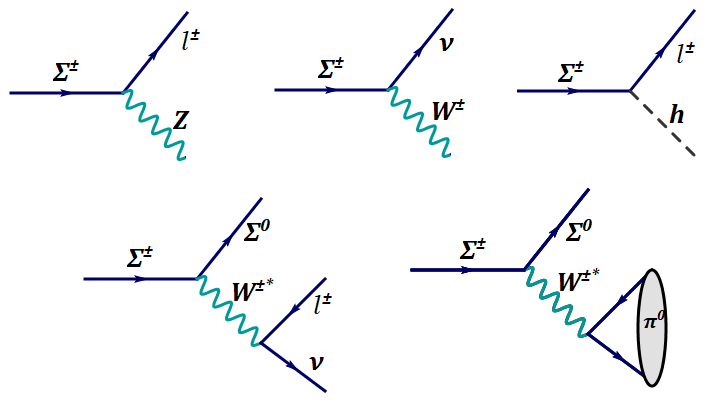} 
\end{center}
 \caption{Representative Feynman diagrams for $\Sigma^{\pm}$ decay. 
 }
\label{fig:decay1}
\end{figure}

We conclude this section by discussing the prospect of probing the type-III seesaw at various current and planned collider experiments. 
After being produced at the collider, the triplet fermion $\Sigma$ decays to the SM particles. 
The type of signal observed at collider depends on the lifetime ($c\tau$) of the charged ($\Sigma^{\pm}$) and the neutral component ($\Sigma^{0}$) of the triplet fermion.  
The Feynman diagrams representing their decays are shown in  Figs. \ref{fig:decay2} and \ref{fig:decay1}. 
Both $\Sigma^{0}$ and $\Sigma^{\pm}$ decay to $W^\pm$/$Z$/$h$ plus SM leptons (hereafter, WZh) while  $\Sigma^{\pm}$ can additionally decay to $\Sigma^{0}$ in association with soft pions/SM leptons. 
The CMS experiment have searched for $\Sigma^{\pm, 0}$ decaying promptly to WZh inside the LHC detector ($c\tau \lesssim 10$ m) and excluded the triplet mass $m_{\Sigma}< 840$ GeV \cite{cms}.  
However, as we will show in the next section, $\Sigma^0$ can have lifetime $c\tau > 10$ m and hence evade prompt-detection at the LHC. 
In this case, the CMS prompt-search bound is not applicable. 
Because the soft pion/SM lepton  accompanying the long-lived $\Sigma_0$ also avoid detection at the LHC, as we have discussed earlier, the LHC and HL-LHC can also search for the disappearing track signature from $\Sigma^{\pm}$ decay. 
Additionally, if the lifetime of $\Sigma^0$ is around 100m, we will show that MATHUSLA detector can discover it. 
On the other hand, the clean (low pile up) environment at the proposed $ep$-collider such as the LHeC and FCC-he enable the  reconstruction of the displaced pion and SM leptons from $\Sigma^{\pm}$ decay. 
Therefore, the LHeC and FCC-he can search for the displaced track signature of a decaying long-lived $\Sigma^{\pm}$. 
Besides these, other interesting scenarios are also possible in the type-III seesaw. 
For example, $\Sigma^{0,\pm}$ could all decay non-promptly inside the LHC detector such that one could simultaneously look for a displaced vertex singature from $\Sigma^{0}$ decay and disappearing track signature from $\Sigma^{\pm}$ decay. 
Because there is no existing collider study of these scenarios, we will not consider them in this work.

\section{Branching ratio and decay width of the fermion triplets}
\label{sec:decay}
The partial decay width of $\Sigma^{0}$ to SM Higgs (h) and gauge bosons ($Z, W^\pm$) are given by \cite{collider1}
\begin{alignat}{3}
\Gamma(\Sigma^{0}_i\to h \nu_\alpha ) &=\Gamma(\Sigma^{0}_i \to h \bar\nu_\alpha ) \;\;&=& \;\; \; \frac{1}{8}\frac{m_{\Sigma_i}^3}{8\pi}
\left(\frac{|R_{\alpha i}|^{2}}{v^2} \right)
\left(1- \frac{m_h^2}{m_{\Sigma_i}^2}\right)^{2}, \nonumber
\\
\Gamma(\Sigma^{0}_i\to Z \nu_\alpha) &=\Gamma(\Sigma^{0}_i\to Z\bar\nu_\alpha) 
\;\;&=&\;\; \;
 \frac{1}{8}\frac{m_{\Sigma_i}^3}{8\pi}
\left(\frac{|R_{\alpha i}|^{2}}{v^2} \right)
\left(1- \frac{M_Z^2}{m_{\Sigma_i}^2}\right)^{2}\left(1+\frac{2 M_Z^2}{m_{\Sigma_i}^2}\right), \nonumber
\\
\Gamma(\Sigma^{0}_i\to W^+ \ell^-_\alpha ) &=\Gamma(\Sigma^{0}_i\to W^- \ell^+_\alpha ) 
\;\;&=&\;\; \;
\frac{1}{4}\frac{m_{\Sigma_i}^3}{8\pi}
\left(\frac{|R_{\alpha i}|^{2}}{v^2} \right)
\left(1- \frac{M_W^2}{m_{\Sigma_i}^2}\right)^{2}
\left(1+\frac{2M_W^2}{m_{\Sigma_i}^2}\right),
\label{eq:PDn}
\end{alignat}
where $\alpha = e,\mu, \tau$ denotes different lepton flavors and 
\bea
R_{\alpha i} =  (m_D)_{\alpha i} (M_\Sigma)^{-1} = U_{\rm{MNS}}^* \sqrt{D_{\nu}} \; O \sqrt{M_\Sigma} (M_\Sigma)^{-1}.  
\label{eq:RMatrix}
\eea
Similarly, the partial decay width of $\Sigma^{\pm}$ to SM Higgs (h) and gauge bosons ($Z, W^\pm$) are given by \cite{collider1}
\bea
\Gamma(\Sigma^{\pm}_i \to h \ell^\pm_\alpha ) &=& \frac{1}{4}\frac{m_{\Sigma_i}^3}{8\pi}\left(\frac{|R_{\alpha i}|^{2}}{v^2} \right)
\left(1- \frac{m_h^2}{m_{\Sigma_i}^2}\right)^{2} ,
\nonumber \\
\Gamma(\Sigma^{\pm}_i \to Z \ell^\pm_\alpha ) &=& \frac{1}{4}\frac{ m_{\Sigma_i}^3}{8\pi} 
\left(\frac{|R_{\alpha i}|^{2}}{v^2} \right)
\left(1- \frac{M_Z^2}{m_{\Sigma_i}^2}\right)^{2}\left(1+\frac{2M_Z^2}{m_{\Sigma_i}^2}\right), 
\nonumber \\
\Gamma(\Sigma^{\pm}_i \to W^\pm  \nubarnu_\alpha ) &=& \frac{1}{2}\frac{ m_{\Sigma_i}^3}{8\pi}
\left(\frac{|R_{\alpha i}|^{2}}{v^2} \right)
\left(1- \frac{2M_W^2}{m_{\Sigma_i}^2}\right)^{2}
\left(1+\frac{2M_W^2}{m_{\Sigma_i}^2}\right). 
\label{eq:PDc1}
\eea 
Because of the mass-splittings, $\Sigma^{\pm}$ can also decay to $\Sigma^0$ and pions or light SM leptons. 
The partial decay widths for these processes are given by \cite{collider1, Cirelli:2005uq}
\bea
\Gamma(\Sigma^{\pm}_i \to \Sigma^{0}_i \pi^\pm ) &=& 
\displaystyle
\frac{2G_{\rm F}^2V_{ud}^2\, \Delta M_i^3 f_\pi^2}{\pi}
\sqrt{1-\frac{m_\pi^2}{\Delta M_i^2}}, 
\nonumber \\
\Gamma(\Sigma^{\pm}_i \to \Sigma^{0}_i e^\pm\nubarnu_e ) &=&\displaystyle
\frac{2G_{\rm F}^2 \,\Delta M_i^5}{15\pi^3},
\nonumber \\
\Gamma(\Sigma^{\pm}_i \to \Sigma^{0}_i \mu^\pm\nubarnu_\mu ) &=&0.12\ \Gamma(\Sigma^{\pm} \to \Sigma^{0} e^\pm\nubarnu_e ). 
\label{eq:PDc2}
\eea
The branching ratios of $\Sigma^{0,\pm}$ decaying to WZh and $\Sigma^0$ are defined as the ratio of the sum of partial decay widths for each decay mode to the total decay width of $\Sigma^{0,\pm}$, $\Gamma_{\rm total}^{\pm,0}$. 
The decay length of $\Sigma^{0,\pm}$ are  given by $c\tau = 1 / \Gamma_{\rm total}^{\pm,0}$. 
From Eqs.~(\ref{eq:PDn}),~(\ref{eq:RMatrix}), (\ref{eq:PDc1}) and (\ref{eq:PDc2}), all the partical decay widths of $\Sigma^{0,\pm}$ are determined by $m_{1,2,3}$, $m_{\Sigma_{1,2,3}}$, and complex angles $\theta_{1,2,3}$. 
The observed light neutrino mass spectrum is uniquely determined by the lightest neutrino mass, particularly, $m_1$ (NH) and $m_3$ (IH). 
For the rest of our analysis, we will also fix  $m_{\Sigma_{2,3 (1,2)}} = 1$ TeV for NH (IH), respectively, which will be justified later. 
Therefore, the branching ratios and decay length of $\Sigma^{0,\pm}_i$ are determined by five free parameters, namely, $m_{1(3)}$ and $m_{\Sigma_{1(3)}}$ for NH (IH), respectively, and $\theta_{1,2,3}$.  
We will consider the two cases, $\theta_{1,2,3} =0$ and $\theta_{1,2,3}\neq 0$, separately.

\begin{figure}[t!]
\begin{center}
\includegraphics[width=0.49\textwidth, height=6.5cm]{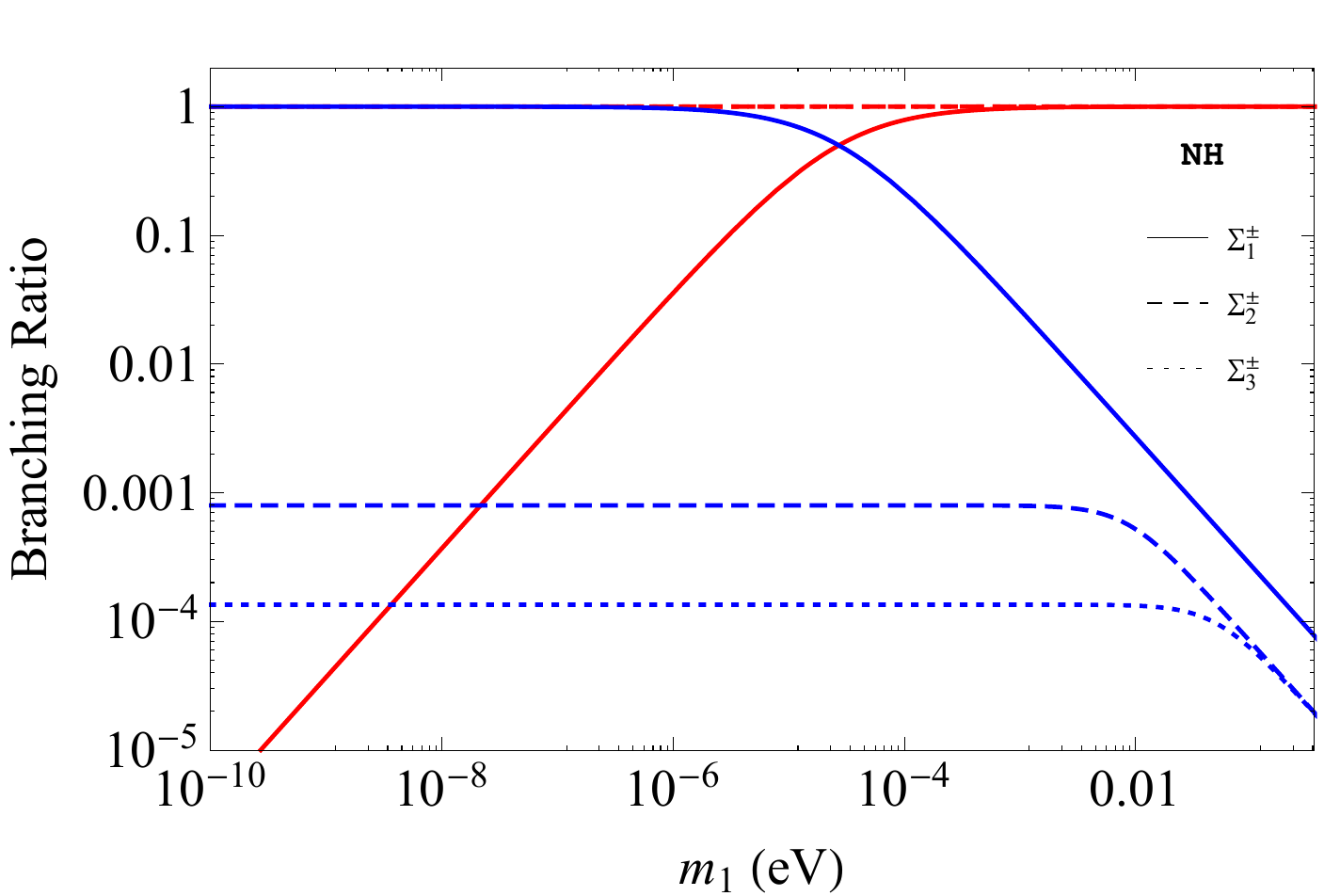}
\includegraphics[width=0.49\textwidth, height=6.5cm]{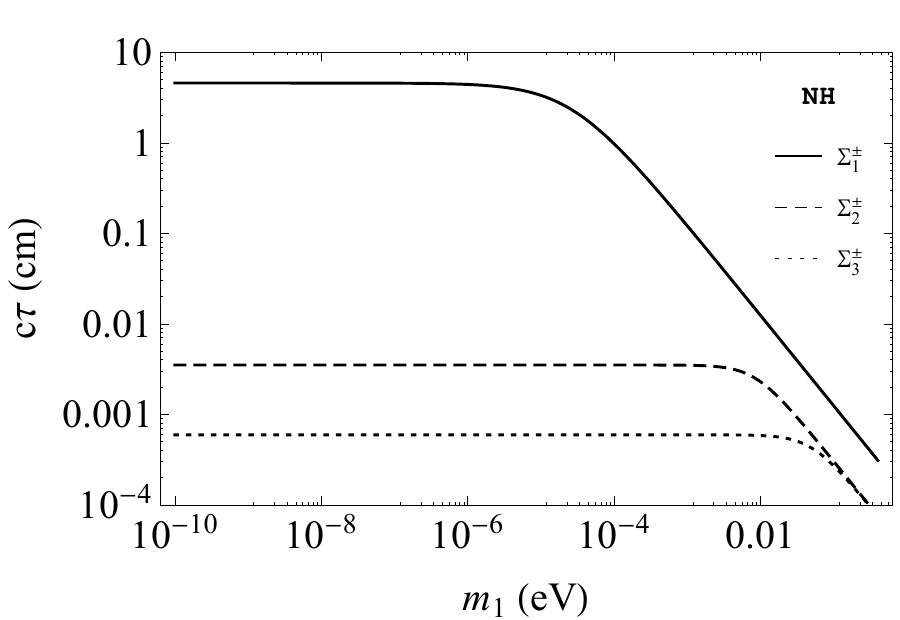} \\
\includegraphics[width=0.49\textwidth, height=6.5cm]{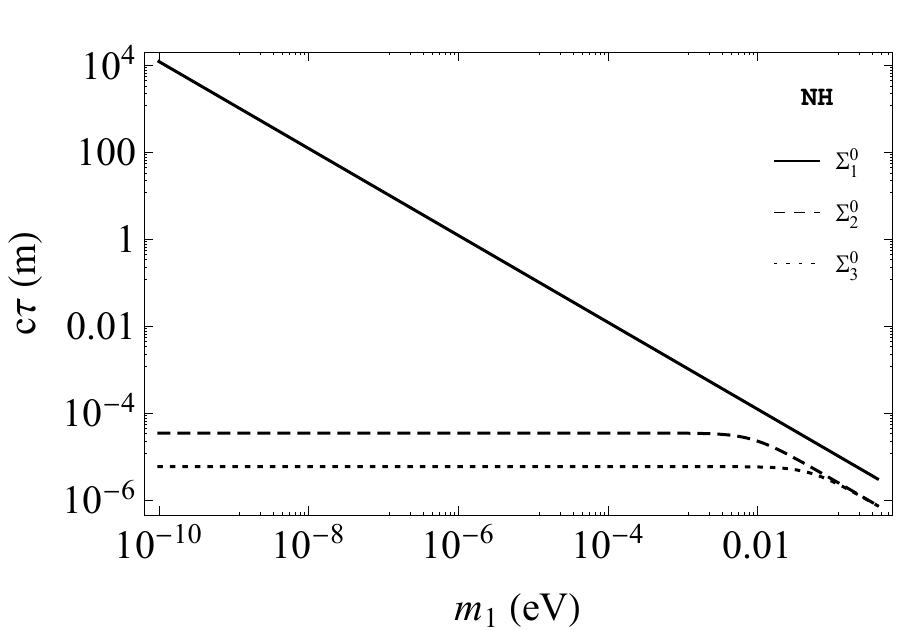}

\end{center}
 \caption{For NH with $\theta_{1,2,3}= 0$,  $m_{\Sigma_1} = 500$ GeV, and $m_{\Sigma_{2,3}} = 1$ TeV, 
the plots shows the branching ratio and decay length of $\Sigma^\pm_i$ in top left and right panels, respectively, and decay length of $\Sigma^0_i$ as a function of lightest observed neutrino mass $m_1$ in the bottom panel. 
The blue (red) lines depict branching ratio for $\Sigma^\pm_i$ decay to $\Sigma^0_i$ (WZh). }
\label{fig:NH}
\end{figure}

In Fig.~\ref{fig:NH}, 
 we consider the case with the $\theta_{1,2,3} =0$ for NH for fixed $m_{\Sigma_1} = 500$ GeV. 
We show the branching ratio for $\Sigma^\pm_i$ as a function of $m_1$ in the top left panel of  Fig.~\ref{fig:NH}. 
The blue (red) lines depict branching ratio for $\Sigma^\pm_i$ decay to $\Sigma^0_i$ (WZh), respectively. 
The solid, dashed and dotted red/blue lines denote $\Sigma^\pm_{1,2,3}$, respectively, as indicated in the figure legend. 
The red dashed and dotted lines overlap indistinguishably and indicate that $\Sigma^\pm_{2,3}$ decay 100\%  to WZh independent of $m_{1}$ values. 
On the other hand, $\Sigma^\pm_{1}$ decays only to WZh for $m_{1} \gg 10^{-4}$ eV. 
For $m_{1} \lesssim 10^{-4}$ eV,  $\Sigma^\pm_{1}$'s branching ratio to WZh drops significantly such that $\Sigma^\pm_{1}$ decays 100\% to $\Sigma^0_{1}$ for $m_{1} \ll 10^{-6}$. 
In the top right panel of Fig.~\ref{fig:NH}, we show the decay length of $\Sigma^\pm_i$ as a function of $m_1$. 
The solid, dashed and dotted lines denote $\Sigma^\pm_{1,2,3}$, respectively. 
The decay length for $\Sigma^\pm_{1,2,3}$ all approach constant values for $m_{1}\to 0$. 
In this limit, the decay length of $\Sigma^\pm_{1}$ approaches a constant value because $\Sigma^\pm_{1}$ decays to $\Sigma^0_{1}$ independently of light neutrino masses. 
On the other hand, $\Sigma^\pm_{2,3}$ decays to WZh for any values of $m_1$. 
The WZh partial decay widths of $\Sigma^\pm_{i}$ in  Eq.~(\ref{eq:PDc1}) are all proportional to $|R_{\alpha i}|^{2}$, which is given by $|R_{\alpha i}|^{2} = m_i /m_{\Sigma_i}$ for $\theta_{1,2,3} = 0$ and is independent of $U_{MNS}$ matrix \cite{dvn1}. 
Hence, the decay length of $\Sigma^\pm_{i}$ is inversely proportional to $m_{i}$. 
For NH ($m_1 < m_2 < m_3$), $m_2  = \sqrt {\Delta m_{21}^2 +m_1^2}$ and $m_3 = \sqrt {m_{2}^2 + \Delta m_{32}^2}$, both of which approach constant values for $m_{1} \to 0$. 
Corresponding to the asymptotic values of $m_{2,3}$, the decay length of $\Sigma^\pm_{2,3}$ also approach constant values in the limit $m_{1} \to 0$.   
For larger values of $m_1^2 \gg \Delta {m_{32}}^2$, both $m_{2,3}$ are proportional to $m_1$, which explains the decreasing behavior of $\Sigma^\pm_{i}$'s decay lengths for large $m_1$ values. 
In the bottom panel of Fig.~\ref{fig:NH}, we show the decay length of $\Sigma^0_i$ as a function of $m_1$. 
The solid, dashed and dotted lines denote $\Sigma^0_{1,2,3}$, respectively. 
The partial decay widths of $\Sigma^0_{i}$ in Eq.~(\ref{eq:PDn}) are all proportional to $|R_{\alpha i}|^{2} = m_i /m_{\Sigma_i}$. 
Therefore, in the limit of $m_{1} \to 0$, the decay length of $\Sigma^0_1$ diverges while the decay length of $\Sigma^0_{2}$ and $\Sigma^0_{3}$ approach constant values corresponding to the asymptotic values of $m_{2,3}$. 
\begin{figure}[t!]
\begin{center}
\includegraphics[width=0.49\textwidth, height=6.5cm]{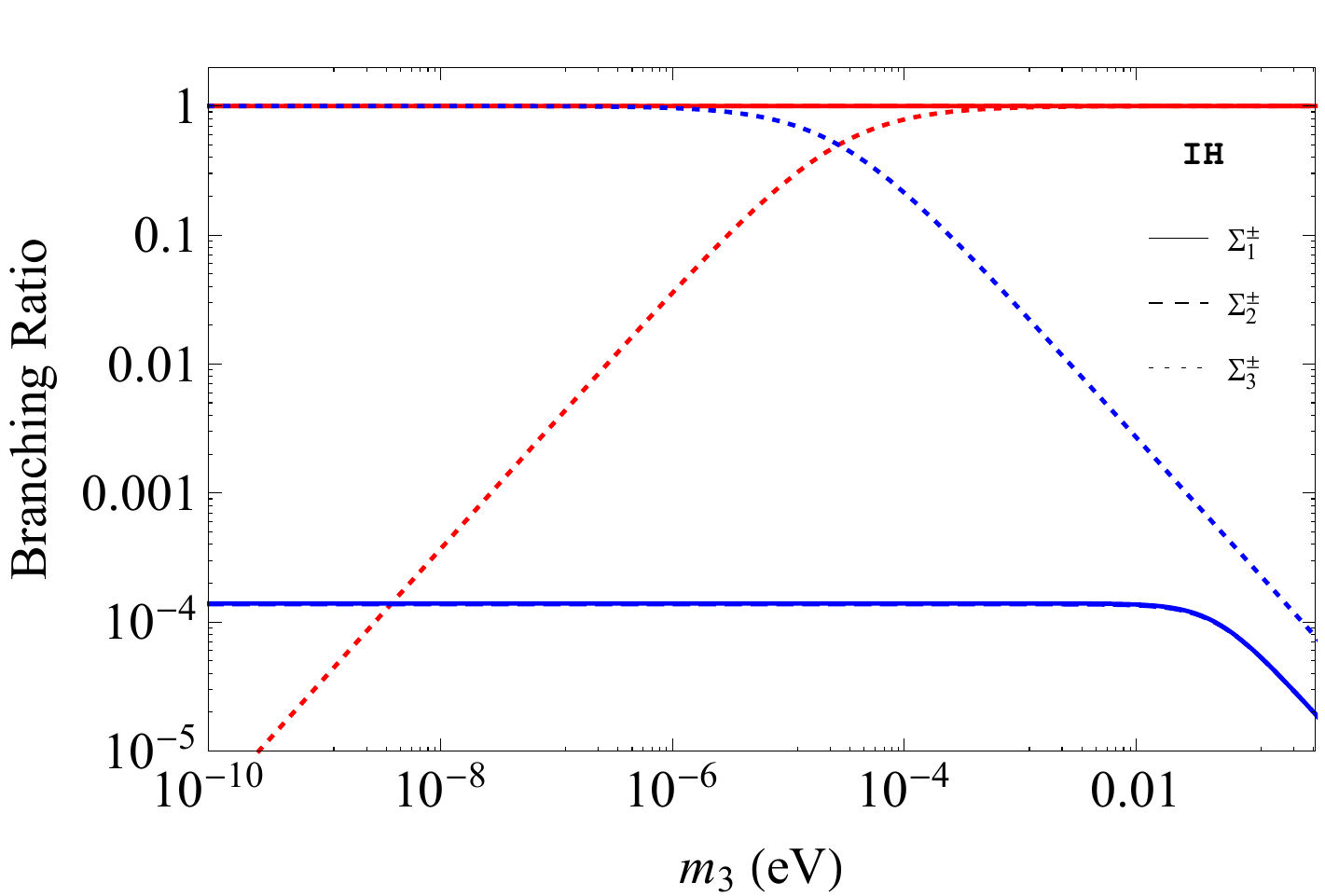} 
\includegraphics[width=0.49\textwidth, height=6.5cm]{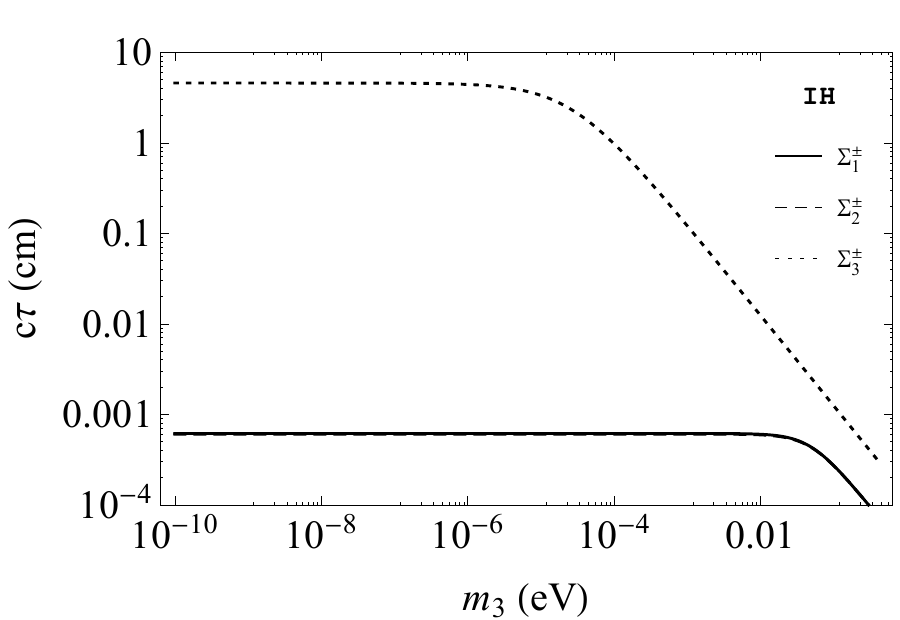} \\
\includegraphics[width=0.49\textwidth, height=6.5cm]{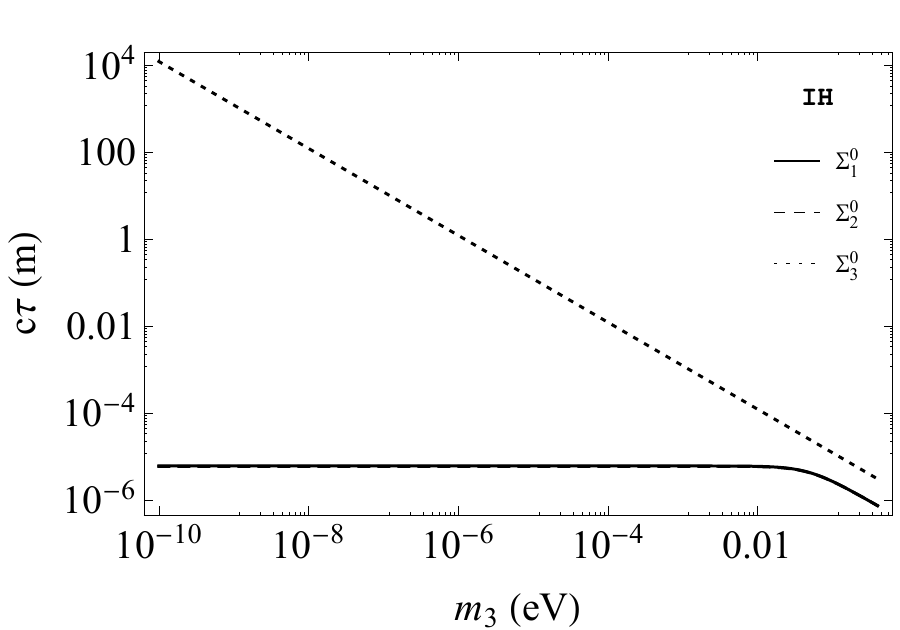} 
\end{center}
 \caption{For NH with $\theta_{1,2,3}= 0$,  $m_{\Sigma_3} = 500$ GeV and $m_{\Sigma_{1,2}} = 1$ TeV, 
the plots show the branching ratio and decay length of $\Sigma^{\pm,0}_i$ as a function of lightest observed neutrino mass $m_1$.  
In the top left and right panels, respectively, and decay length of $\Sigma^0_i$ as a function of lightest observed neutrino mass $m_1$ in the bottom panel. 
The blue (red) lines depict branching ratio for $\Sigma^\pm_i$ decay to $\Sigma^0_i$ (WZh). }
\label{fig:IH}
\end{figure}

\begin{figure}[t!]
\begin{center}
\includegraphics[width=0.45\textwidth, height=6cm]{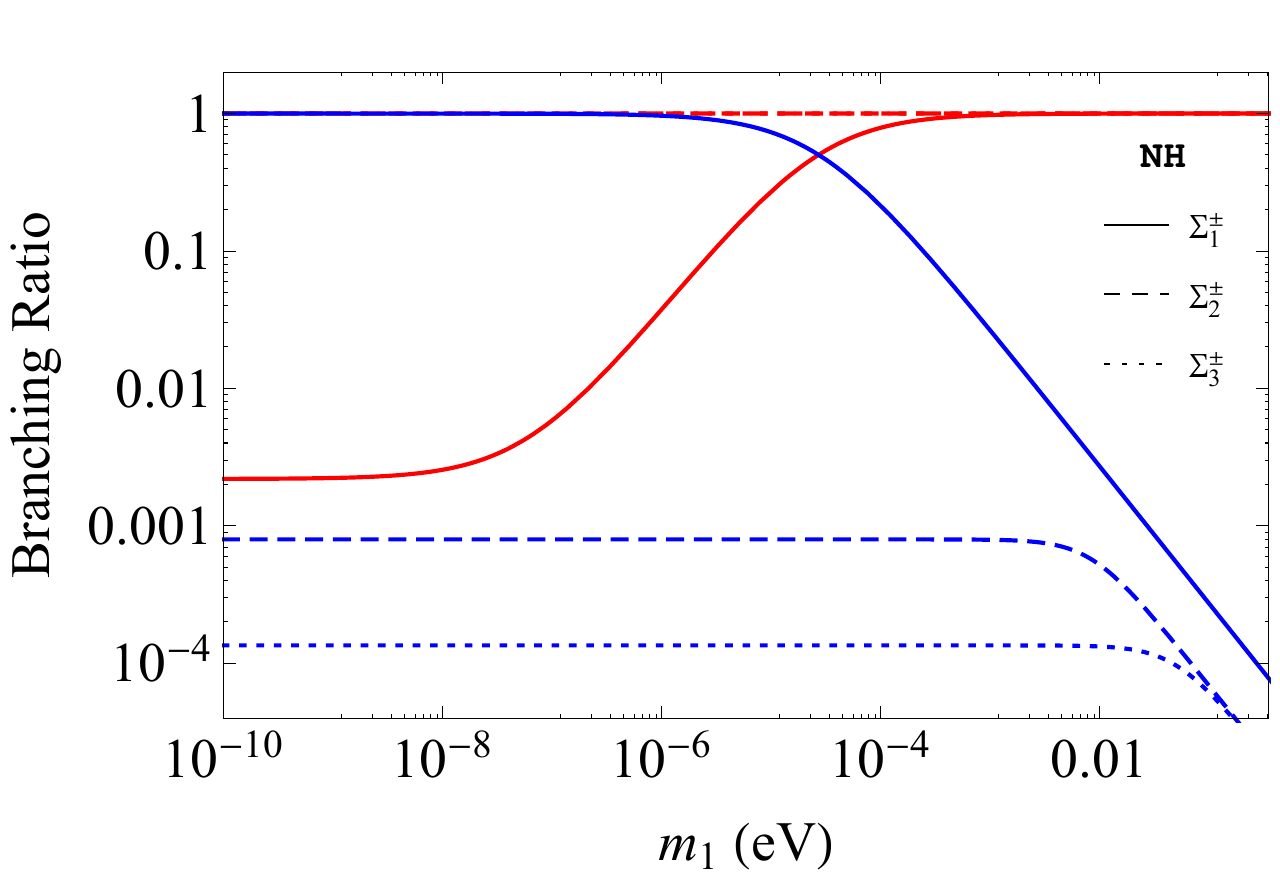} 
\includegraphics[width=0.45\textwidth, height=6cm]{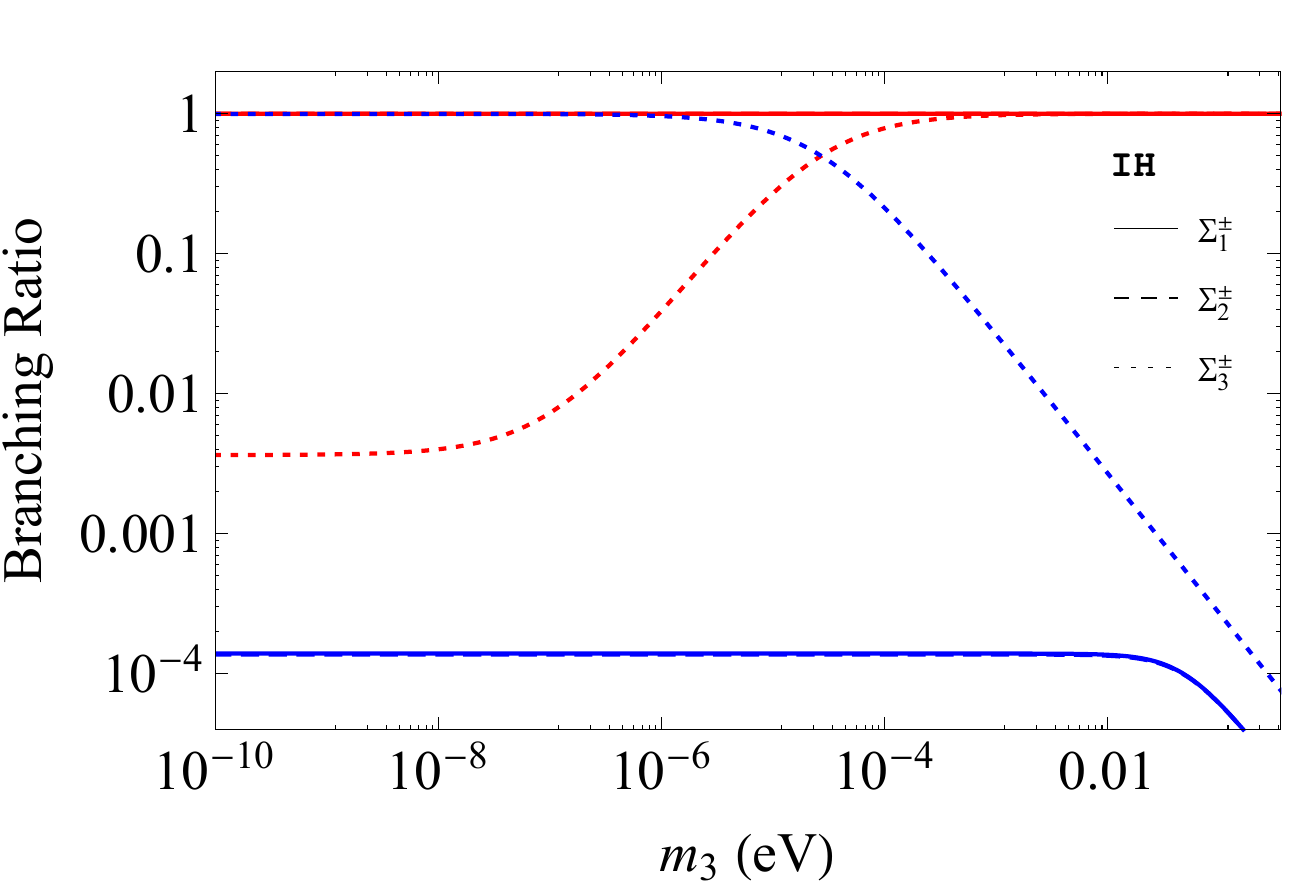} \\
\includegraphics[width=0.45\textwidth, height=6cm]{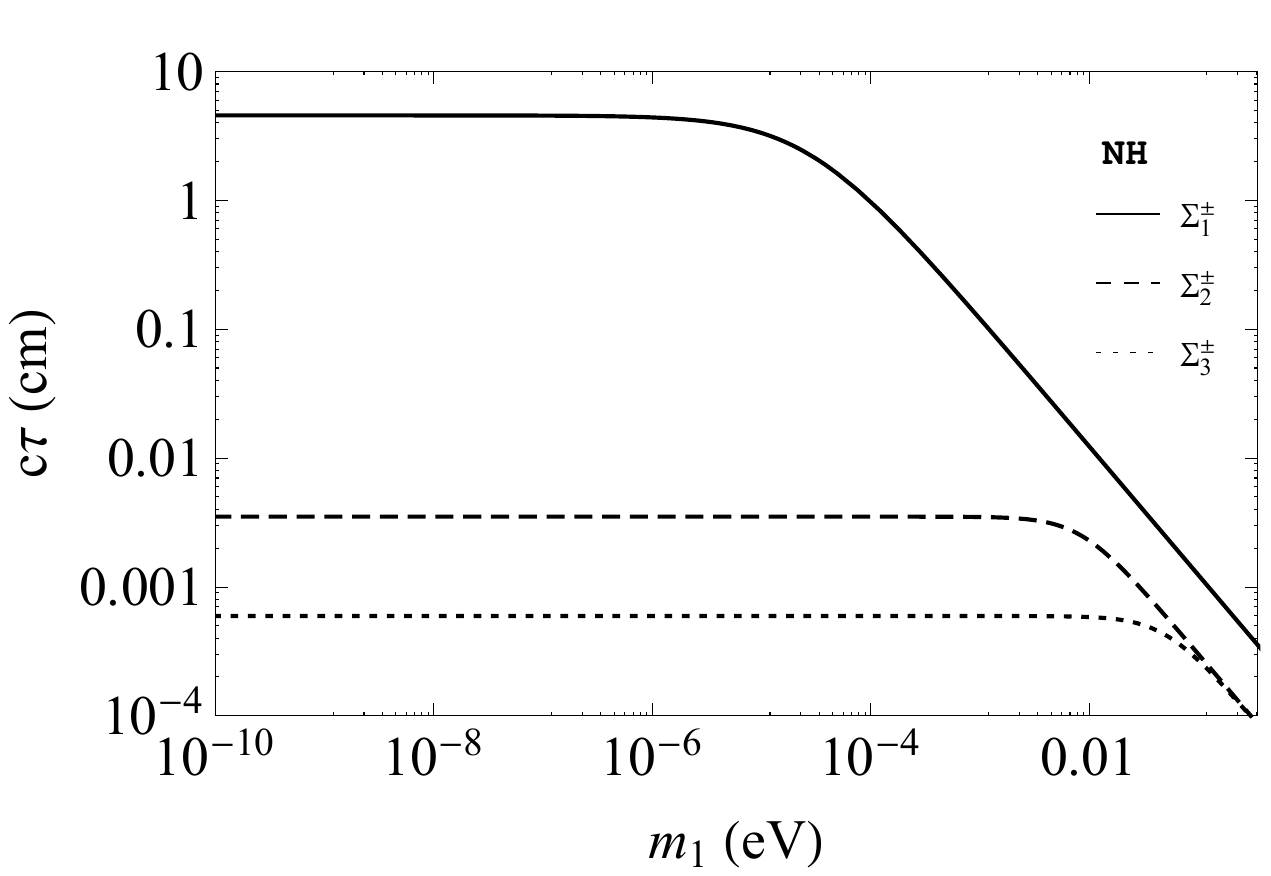}
\includegraphics[width=0.45\textwidth, height=6cm]{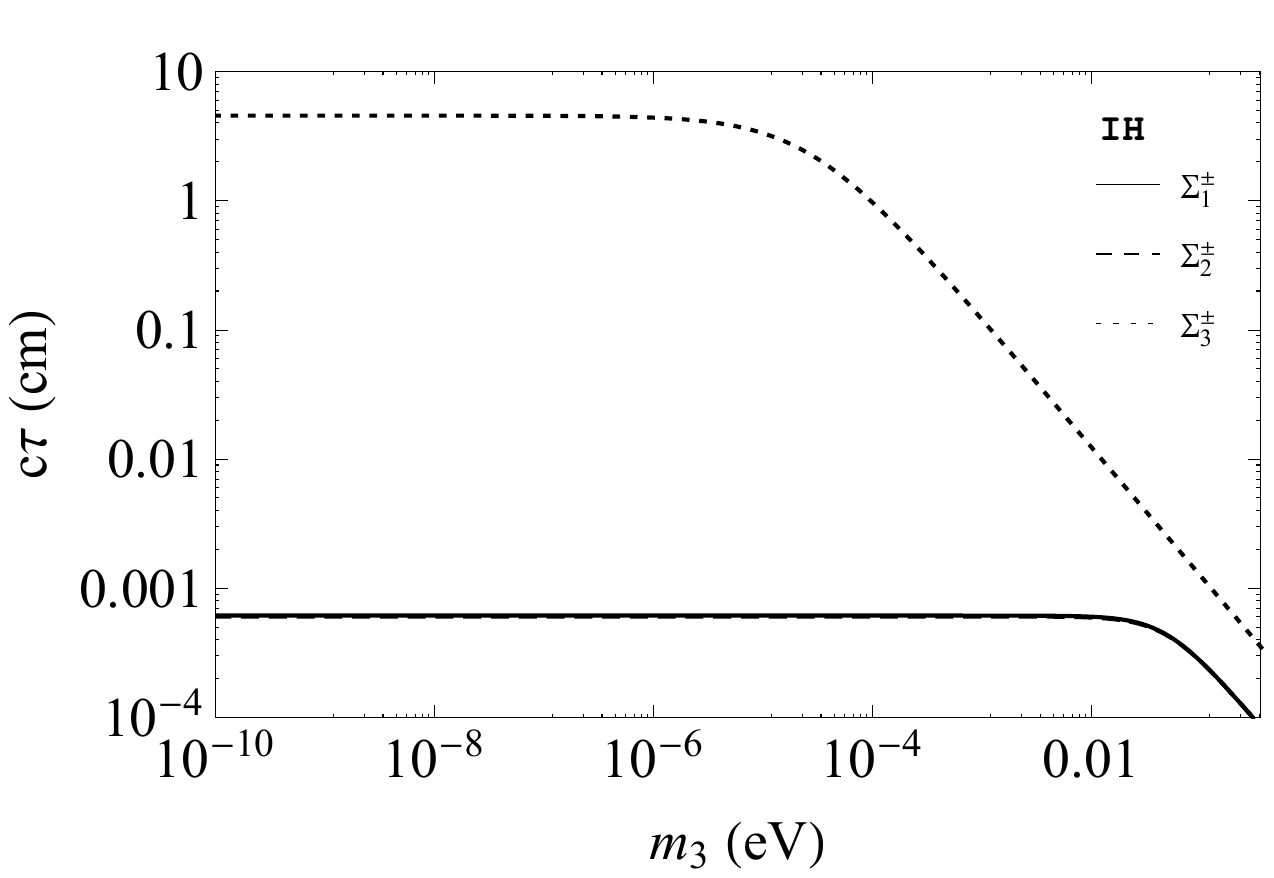}\\
\includegraphics[width=0.45\textwidth, height=6cm]{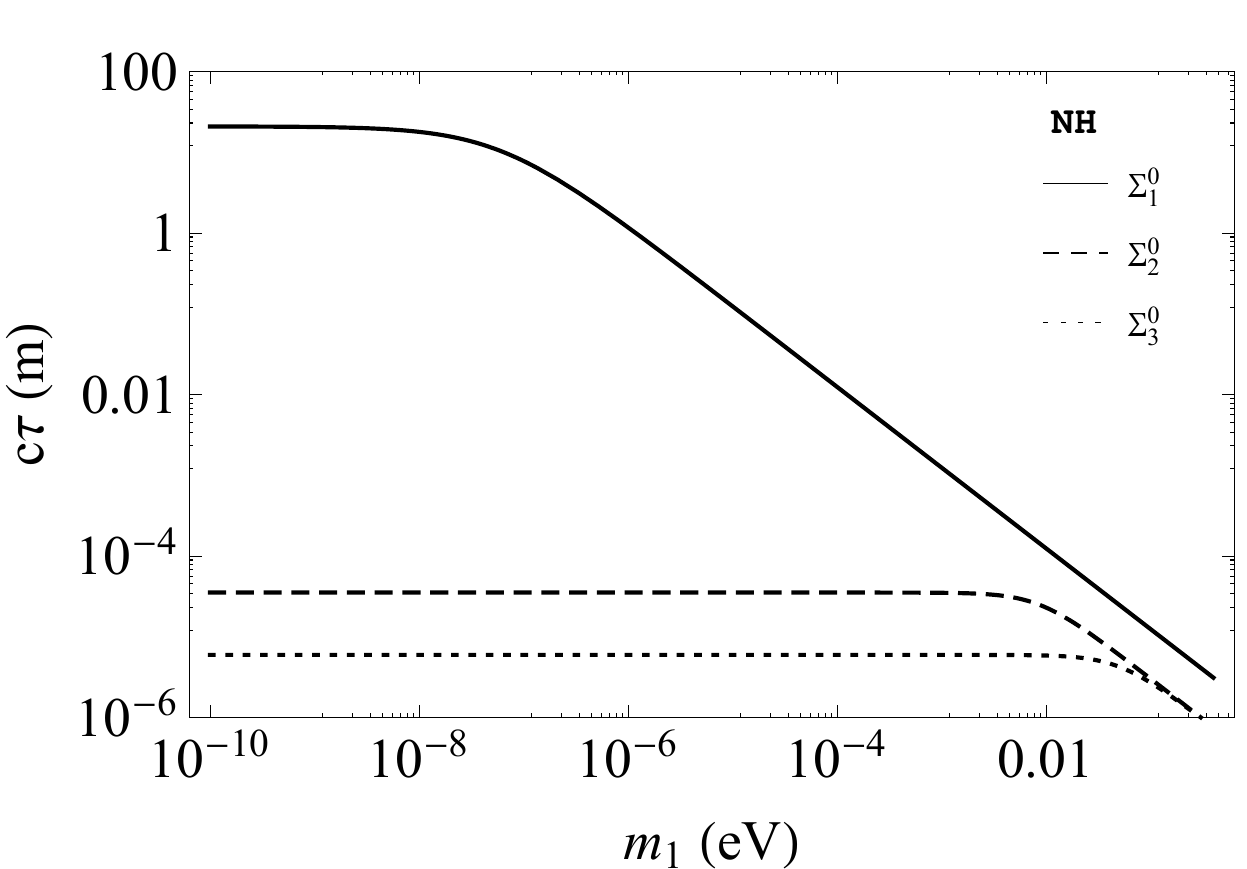}
\includegraphics[width=0.45\textwidth, height=6cm]{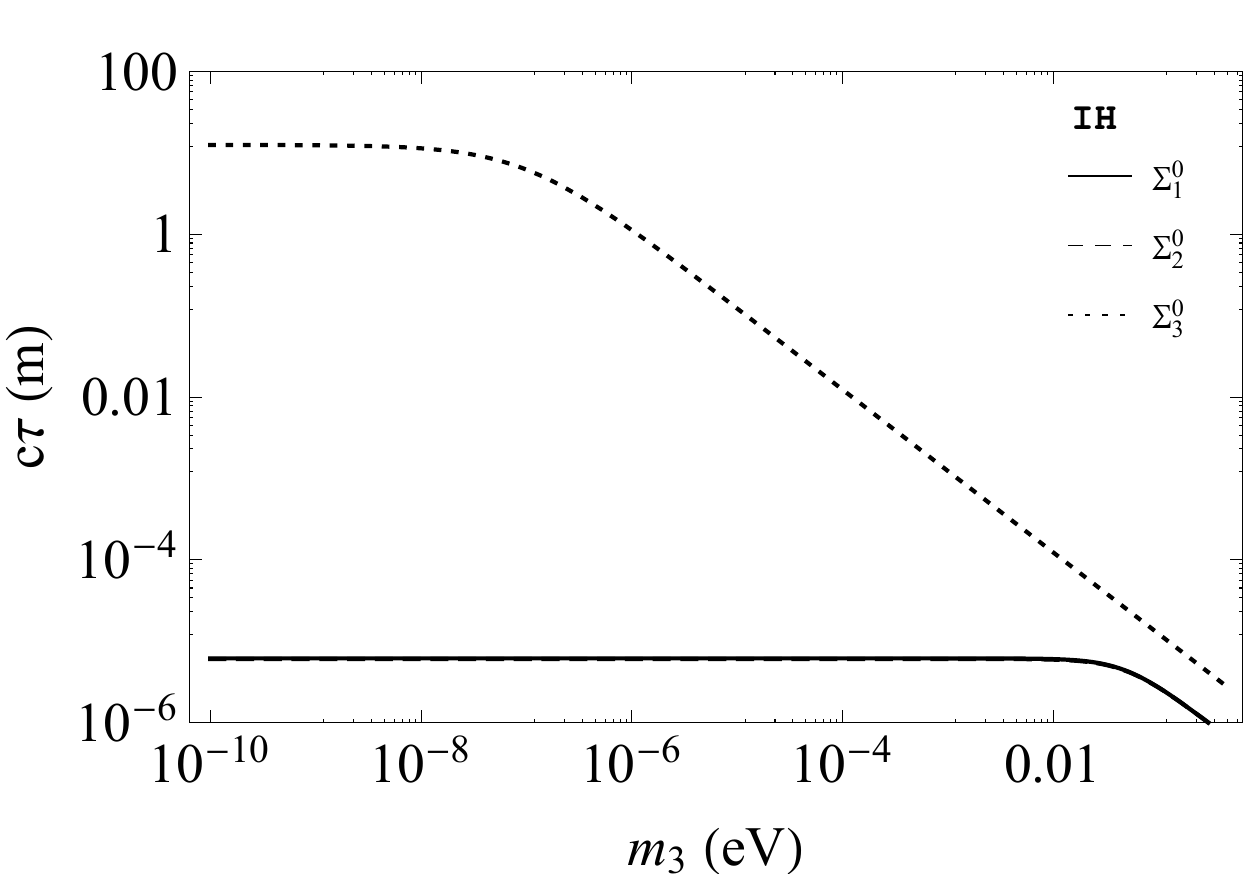}
\end{center}
\caption{For IH with $m_{\Sigma_1} = 500$ GeV, $m_{\Sigma_{2,3}} = 1$ TeV, $x_{1,2,3} =0$ and $y_{1,2,3} =10^{-3}$, 
the left panels show the branching ratio (top), decay length of $\Sigma^\pm_i$ (middle) and $\Sigma^0_i$ (bottom) as a function of the lightest observed neutrino mass ($m_3$). All the line codings are the same as Fig.~\ref{fig:NH}. The right panels show the same for the IH case, with $m_{\Sigma_3} = 500$ GeV, $m_{\Sigma_{1,2}} = 1$ TeV, $x_{1,2,3} =0$ and $y_{1,2,3} =10^{-3}$ as a function of $m_3$ and all the line codings are the same as Fig.~\ref{fig:IH}.}
\label{fig:COM}
\end{figure}
In Fig.~\ref{fig:IH}, 
 we consider the IH ($m_3 < m_1 < m_2$) case with $\theta_{1,2,3} =0$, $m_{\Sigma_3} = 500$ GeV and $m_{\Sigma_{1,2}} = 1$ TeV. 
The line codings of all the figures are the same as Fig.~\ref{fig:NH}. 
The branching ratio and decay length are determined as a function of $m_3$, the lightest mass eigenvalue among the observed neutrinos for the IH case.  
The top left panel shows that $\Sigma^\pm_{1,2}$, depicted as indistinguishably overlapped solid and dashed red lines, decay 100\%  to WZh independently of $m_3$ values. 
It also shows that $\Sigma^\pm_{3}$ decays 100\% to $\Sigma^{0}_3$ (WZh) in the limit $m_3 \to 0$ ($m_3 \to 0.3$ eV). 
The discussion to understand the behavior of the decay width of $\Sigma^{\pm,0}_{i}$ is very much analogous to that of the NH case. 
For example, the longest decay length is realized for $\Sigma^{\pm}_{3}$ and $\Sigma^{0}_{3}$ corresponding to the lightest neutrino mass in the IH case.  
The decay length of $\Sigma^{\pm}_{3}$,  depicted as indistinguishably overlapped solid and dashed  lines, approaches a constant value because the corresponding partial decay widths are independent of neutrino masses while the decay length of $\Sigma^{0}_{3}$ diverges because it decays dominantly to WZh with the partial decay width proportional to $m_3$. 
Similarly, $\Sigma^\pm_{1,2}$ decay is also dominated decay to WZh and is proportional to $m_{1,2}$, respectively. Therefore, the decay length $\Sigma^\pm_{1,2}$ approach the same  constant value in the limit $m_3\to 0$ because for the IH, $m_2 \to \sqrt {- \Delta m_{32}^2}$ and $m_1 \to \sqrt {- \Delta m_{32}^2- \Delta m_{21}^2} \simeq \sqrt {- \Delta m_{32}^2}$.

Next, we consider the case $\theta_{1,2,3} \neq 0$. 
Let us parameterize $\theta_1 = x_1 + i y_1$, $\theta_2 = x_2 + i y_2$, and $\theta_3 = x_3 + i y_3$. 
In the following we set $x_{1,2,3} = 0$ for simplicity. 
In this case, $\cos(x_i + i y_i) \simeq \sin(x_i + i y_i) \simeq e^{y_i}$ for $y_i\gtrsim 1$. 
Hence, $|R_{\alpha i}|^{2}$ can be exponentially enhanced for nonzero values of $y_i$. 
To see this effect, we consider $\Sigma^{\pm}_i$ decay in Fig.~\ref{fig:COM} with $m_{\Sigma_1 (3)} = 500$ GeV, $m_{\Sigma_{2,3}(1,2)} = 1$ TeV, respectively, and the angles  $x_{1,2,3} =0$ and $y_{1,2,3} = 10^{-3}$ ($10^{-3}$) for NH (IH).
In this figure, for both NH and IH, curves corresponding to $\Sigma^{\pm/0}_{1,2,3}$ are depicted by solid, dashed and dotted lines.  
Also for both NH and IH, the red (blue) colored lines denote the branching ratio to WZh ($\Sigma_0$). 
The $y_i$ values for the NH (IH) are chosen specifically to suppress the branching ratio of $\Sigma^\pm_1$ ($\Sigma^\pm_3$) to WZh and also realize the lifetime for $\Sigma^0$ associated with the lightest observed neutrino, namely, $\Sigma^0_1$ ($\Sigma^0_3$) to be greater than 10 m. 
\begin{figure}[t!]
\begin{center}
\includegraphics[width=0.45\textwidth, height=6cm]{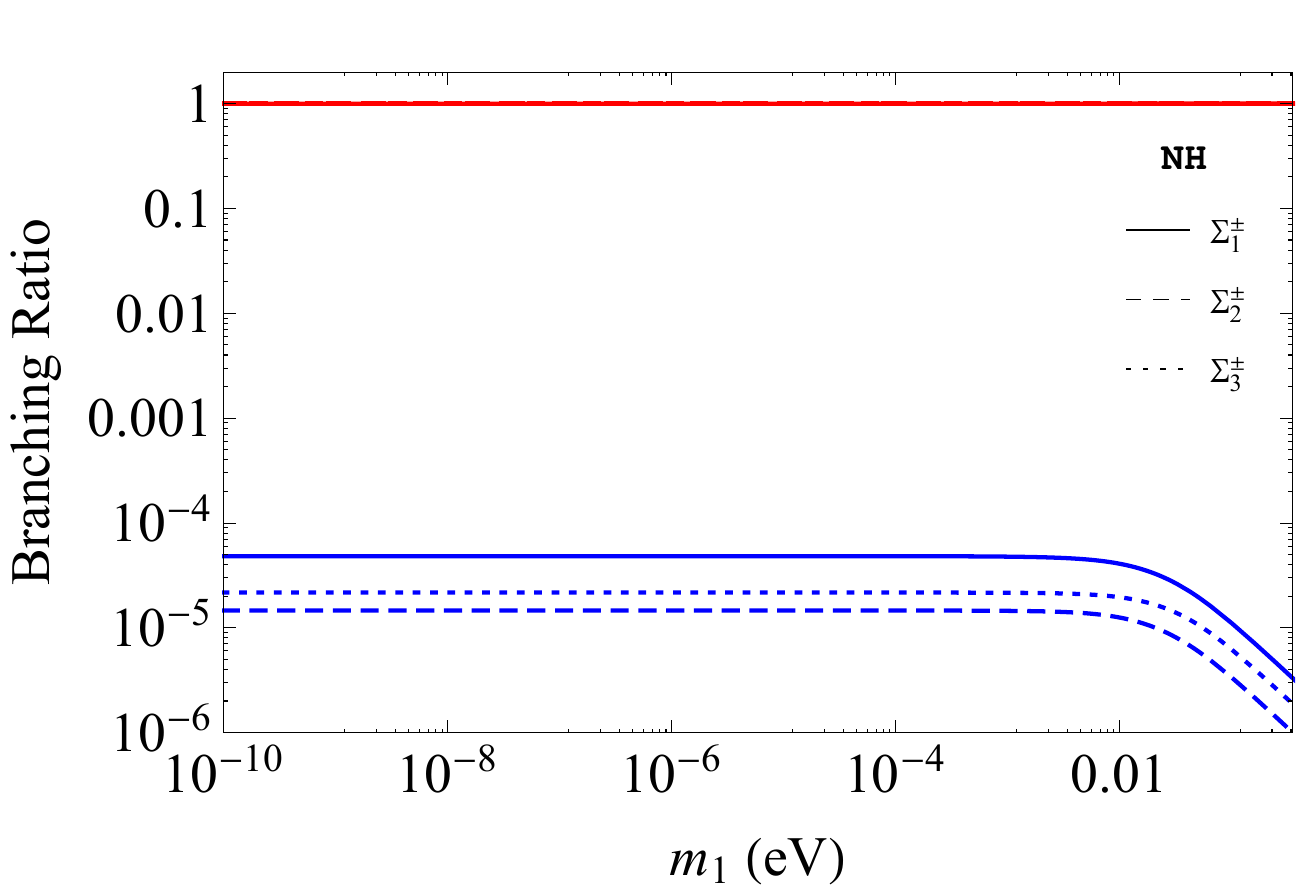} 
\includegraphics[width=0.45\textwidth, height=6cm]{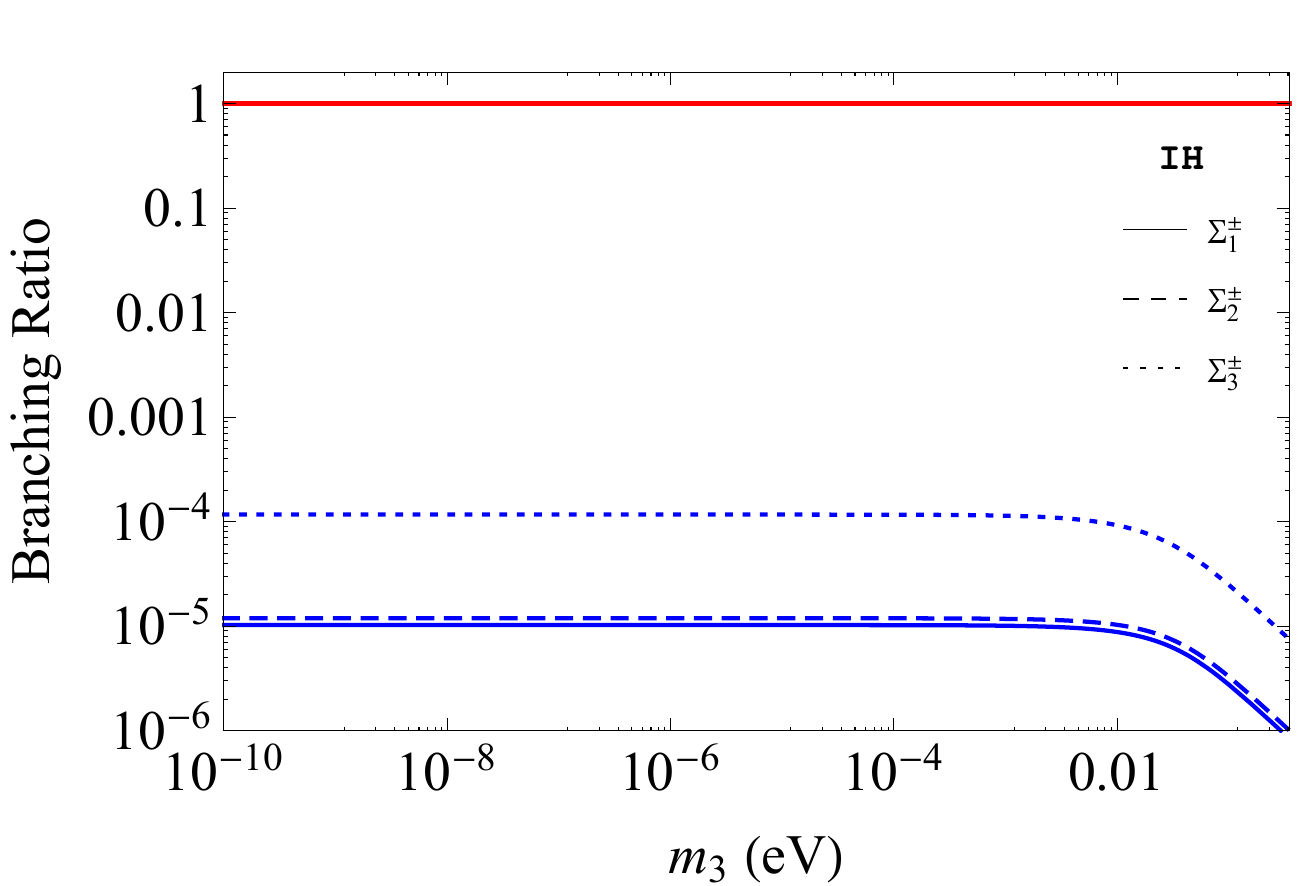} \\
\includegraphics[width=0.45\textwidth, height=6cm]{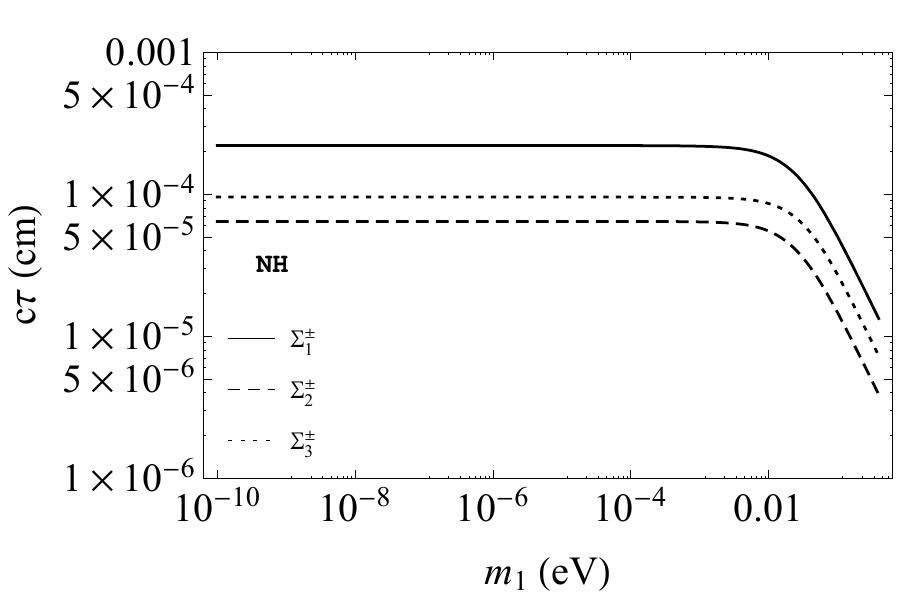}
\includegraphics[width=0.45\textwidth, height=6cm]{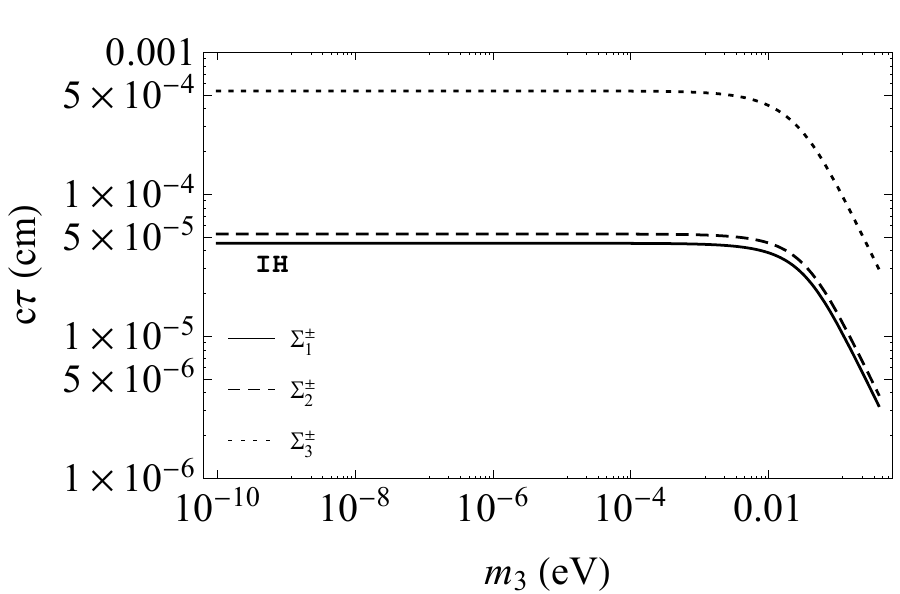}
\end{center}
\caption{For NH with $m_{\Sigma_1} = 500$ GeV, $m_{\Sigma_{2,3}} = 1$ TeV, $x_{1,2,3} =0$ and $y_{1,2,3} =1$, 
the left panels show the branching ratio (top) and decay length of $\Sigma^\pm_i$ (middle) as a function of the lightest observed neutrino mass ($m_1$). 
All the line codings are the same as Fig.~\ref{fig:NH}. The right panels show the same for the IH case, with $m_{\Sigma_3} = 500$ GeV, $m_{\Sigma_{1,2}} = 1$ TeV, $x_{1,2,3} =0$ and $y_{1,2,3} =10^{-3}$ as a function of $m_3$ and all the line codings are the same as Fig.~\ref{fig:IH}.  For both NH and IH, the decay length of the neutral components is the same as the decay length of its charged partner. (See text for details) }
\label{fig:COM1}
\end{figure}

For completeness, let us also consider the case with even larger $y_i$ values. 
We show our results in Fig.~\ref{fig:COM1} for $m_{\Sigma_{1 (3)}} = 500$ GeV, $m_{\Sigma_{{2,3}(1,2)}} = 1$ TeV for NH (IH), respectively, with  $x_{1,2,3} =0$ and $y_{1,2,3} = 1$. 
The line codings for all the figures are the same as in Fig.~\ref{fig:COM}. 
The branching ratio plots (top panel) shows that the decay of all triplets $\Sigma^{0,\pm}_{1,2,3}$, for both NH and IH, are dominated by their decay to WZh. 
In this case, the total decay width for both charged and neutral triplets in each $\Sigma_{i}$ are identical as shown in Eqs.~(\ref{eq:PDn}) and (\ref{eq:PDc1}). 
Hence, the the decay length of $\Sigma^{0}_{i}$ is same as that of $\Sigma^{\pm}_{i}$ shown in the bottom panels of Fig.~\ref{fig:COM1}. 
This shows that their decays are effectively prompt and therefore the choice of $m_{\Sigma_{1(3)}} = 500$ GeV for NH (IH) is inconsistent with the the lower bound on promptly decaying triplet fermions of the type-III seesaw of 840 GeV.

In general, we find that all the triplet fermions, except for the one associated with the lightest observed neutrino (${\Sigma_{1(3)}}$ for NH (IH), respectively), always decays promptly for all values of $y_i\geq 0$. 
See, for example, Figs.~\ref{fig:NH}, \ref{fig:IH}, \ref{fig:COM} and \ref{fig:COM1}. 
Therefore, for NH (IH), ${\Sigma_{1(3)}}$  masses are subject to the CMS bound, namely, $m_{\Sigma_{1}} > 840$ GeV.  
For example, this excludes the IH scenario with $m_{\Sigma_{1}} = 500$ GeV and  $m_{\Sigma_{2,3}} = 1$ TeV.

\section{Prospect to search for fermion triplets at current/future colliders} \label{sec:dv}
In the previous section we have shown that the neutral lepton associated with the lightest observed neutrino, particularly $\Sigma^0_{1 (3)}$ for NH (IH), respectively, can have lifetime $c\tau \gg 10$ m and evade prompt detection at collider while their respective charged partners $\Sigma^\pm_{1 (3)}$ have a much shorter lifetime and decay inside the detector. 
For the remainder of this paper, we will refer both of them as ${\tilde \Sigma}$ and their masses as $m_{\tilde \Sigma}$, which is  effectively the same for both the charged and the neutral components of ${\tilde \Sigma}$. 
In this section, we will study the prospect to  search for the disappearing track signature from ${\tilde \Sigma}^\pm$ decaying inside the LHC and the displaced vertex signature from ${\tilde \Sigma}^0$ decaying inside the MATHUSLA detector. 
We will also study the prospect to search for the displaced vertex signature from ${\tilde \Sigma}^\pm$ decaying inside the LHeC and FCC-he.

\subsection{Disappearing track searches for $\Sigma^\pm$ at the LHC and HL-LHC}
Although there is no dedicated LHC analysis of the disappearing track signature for the type-III seesaw, we can apply the LHC search result for disappearing track signal from Higgsino decay \cite{Sirunyan:2018ldc}. 
This is because the production processes of the charged (neutral) components of Higgsino is the same as that of ${\tilde \Sigma}^\pm$ (${\tilde \Sigma}^0$) components of the type-III seesaw triplets. 
Before we show our results let us highlight the key similarities and differences between Higgsino and type-III seesaw triplets. 
Although the production processes of ${\tilde \Sigma}^{\pm,0}$ are the same as the Higgsino, their production rates are different due to their $SU(2)_L$ representations.  
Unlike the triplets, the Higgsino (charged and neutral) do not decay to WZh at the tree-level. 
On the other hand, the charged Higgsino decays to neutral Higgsino plus soft pions/SM leptons is the same as ${\tilde \Sigma}^\pm$ decay. 
However, the decay length of a charged Higgsino is essentially a free parameter because the mass-splitting between the charged and neutral Higgsinos, which determines the decay length, is a free parameter in the supersymmetric SM. 
In the type-III seesaw case, the decay length depends on the mass-splitting, which is fixed to be around the pion mass, and neutrino oscillation parameters as we have shown in the previous section.

\begin{figure}[t!]
\begin{center}
\includegraphics[width=0.49\textwidth, height=6.5cm]{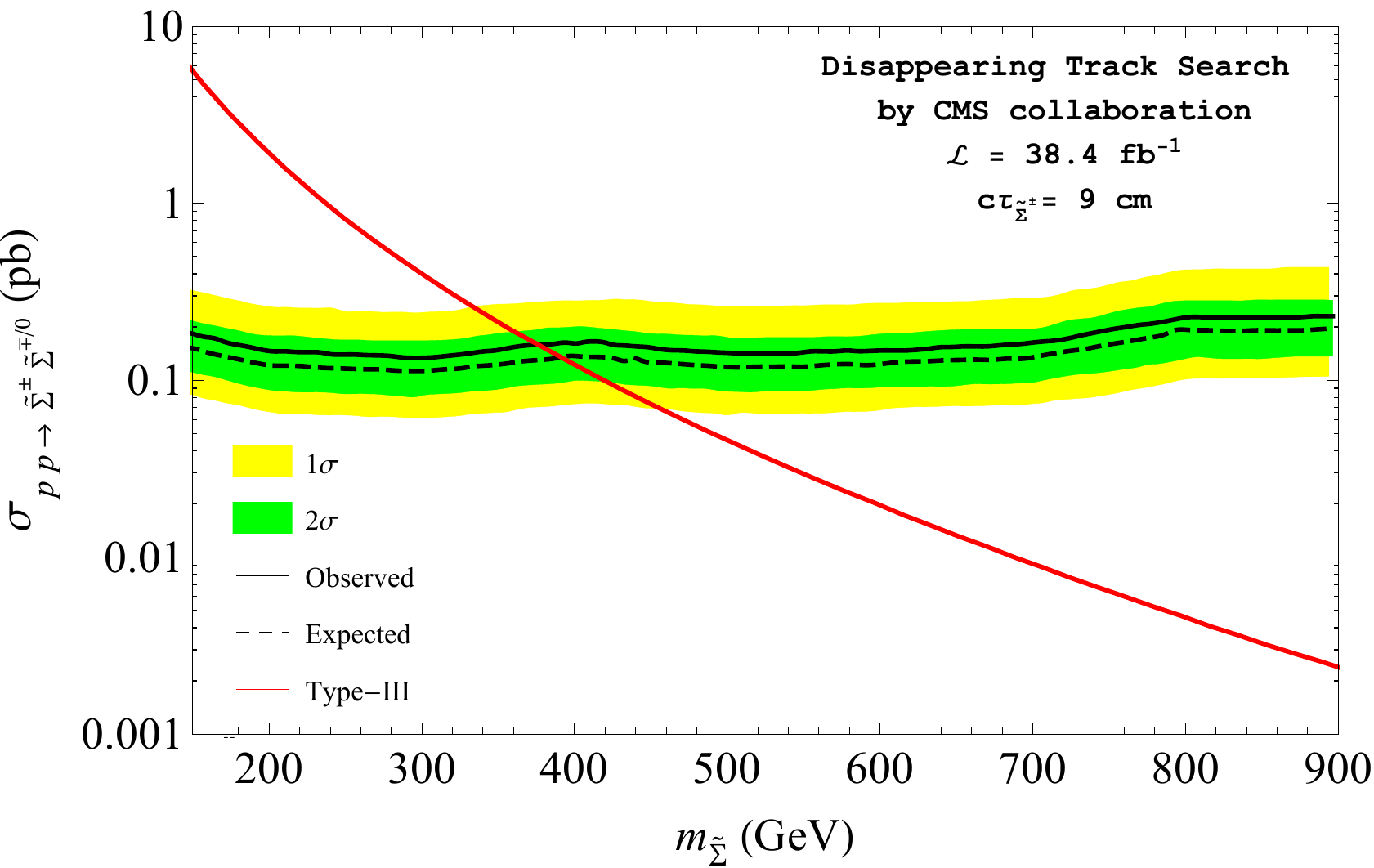}
\includegraphics[width=0.49\textwidth, height=6.5cm]{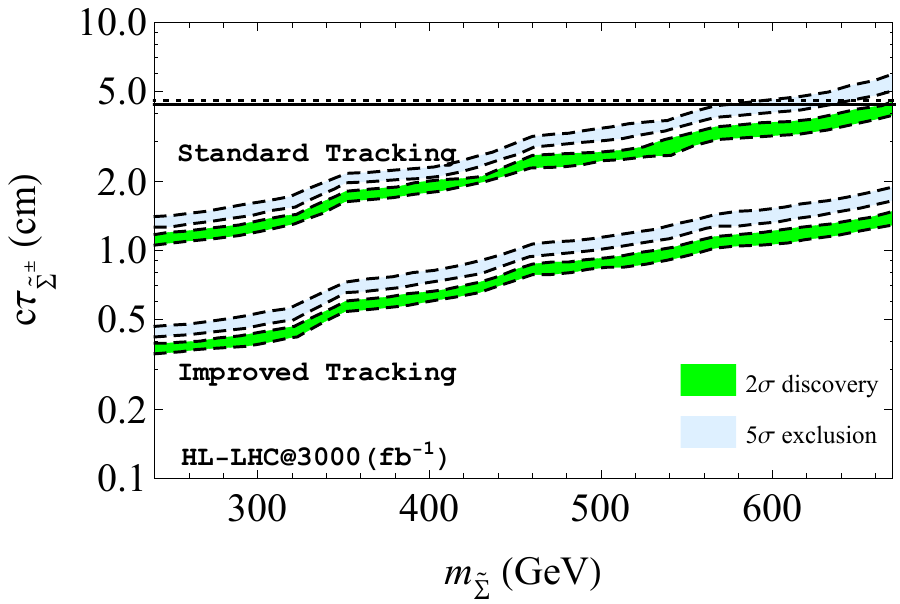}
\end{center}
 \caption{The left panel shows the CMS bound on the production cross section of ${\tilde \Sigma}^{\pm}$ with lifetime $c\tau = 9$ cm as a function of its mass $m_{\tilde \Sigma}$.  
The solid (dashed) black horizontal lines are the observed (expected) limits on the cross section while the green (yellow) band are expected 2(3)-$\sigma$ upper limit on the production cross section. 
The red diagonal line is our result for the total production cross section of the ${\tilde \Sigma}^{\pm}$ at the LHC in Fig.~\ref{fig:epcs}.  
The right panel shows the prospect to search for ${\tilde \Sigma}^{\pm}$ disappearing track signature at high luminosity (HL)-LHC with $3000$ fb$^{-1}$ luminosity. 
The solid (dotted) horizontal black lines are the decay lengths of ${\tilde \Sigma^\pm}$ corresponding to NH (IH), respectively, for the benchmark values $\theta_{1,2,3} =0$,  $m_{{\tilde \Sigma}_1} = 500$ GeV,  $m_{\Sigma_{2,3}} (m_{\Sigma_{1,2}}) = 1$ TeV and $m_{1(3)}  = 10^{-9}$ eV.  
}
\label{fig:distrack}
\end{figure}

The result by the CMS collaboration in Ref.~\cite{Sirunyan:2018ldc} is expressed as a bound on the product of the total production cross section (including both charged and neutral Higgsino) and the branching fraction of the charged Higgsino decaying to neutral Higgsino plus a pion which is assumed to be 100\%. 
This branching ratio is realized in the type-III seesaw in the benchmark scenarios shown in  Figs.~\ref{fig:NH}, \ref{fig:IH} and \ref{fig:COM} in the limit $m_{1 (3)} \to 0$ for NH (IH), respectively. 
In this limit, we can identify the bound in Ref.~\cite{Sirunyan:2018ldc} as a upper bound bound on the total production cross section of ${\tilde \Sigma}^{0,\pm}$, which includes ${\tilde \Sigma}^0 {\tilde \Sigma}^\pm$ and ${\tilde \Sigma}^\mp {\tilde \Sigma}^\pm$ final states. 
We show our result in the left panel of Fig.~\ref{fig:distrack}, where the diagonal red line denotes the total production cross section for ${\tilde \Sigma}^{\pm,0}$ as a function of $m_{\tilde \Sigma}$. 
This red line in Fig.~\ref{fig:distrack} is taken from the left panel of Fig.~\ref{fig:epcs}. 
The solid (dashed) black horizontal lines are the observed (expected)  cross section limits from Ref.~\cite{Sirunyan:2018ldc} while the green (yellow) band are expected 2(3)-$\sigma$ upper limit on the production cross section. 
As indicated in the Fig.~\ref{fig:distrack}, this bound was obtained for Higgsino lifetime is fixed to be 9 cm, which is longer than $c\tau_{{\tilde \Sigma}^{\pm}} \simeq 5$ cm \footnote{This result was obtained by using the mass-splitting formula that only included one-loop contributions. Taking into account the two-loop contribution, the decay length of ${\tilde \Sigma}^{\pm}$ can increase by as much as $10$\%  \cite{Yamada:2009ve}.} for ${\tilde \Sigma}^{\pm}$ in the limit $m_{1 (3)} \to 0$ for NH (IH), respectively.  
According to Ref.~\cite{Sirunyan:2018ldc}, the bound on the cross section becomes weaker for shorter Higgsino lifetime. 
Hence, our choice for $m_{\tilde \Sigma} = 500 $ GeV is consistent with the lower bound on the triplet mass obtained from  Fig.~\ref{fig:distrack}, namely, $m_{\tilde \Sigma} \gtrsim 380$ GeV.  

Let us now consider a disappearing track search prospect for ${\tilde \Sigma}^{\pm}$ at the HL-LHC by interpreting  the charged Higgsino disappearing track search prospect examined in Ref.~\cite{Mahbubani:2017gjh}.  
To do so we also need to take into account the difference in the production rate of Higgsino and ${\tilde \Sigma}$. 
For a fixed value of mass, the difference in production rate is only from the difference in the $SU(2)_L$ representations of Higgsino and ${\tilde \Sigma}$, namely, Higgsino belong to the $SU(2)_L$ doublet while ${\tilde \Sigma}$ is $SU(2)_L$ triplet.   
The Higgsino disappearing track search reach is presented as a plot for Higgsino lifetime as a function of its mass in Fig.~9a of Ref.~\cite{Mahbubani:2017gjh}. 
To interpret this result, we combine the one-to-one correspondence between the production rate of Higssino and ${\tilde \Sigma}$ (for a fixed value of mass) with the one-to-one correspondence between the Higgsino mass, its production rate and lifetime in Ref.~\cite{Mahbubani:2017gjh}. 
We show our result in the right panel of Fig.~\ref{fig:distrack}. 
We also show the decay length of ${\tilde \Sigma^\pm}$ corresponding to NH (IH), which are depicted as the solid (dotted) horizontal black lines for 
the benchmark values $\theta_{1,2,3} =0$,  $m_{{\tilde \Sigma}_1} = 500$ GeV,  $m_{\Sigma_{2,3}} (m_{\Sigma_{1,2}}) = 1$ TeV and $m_{1(3)}  = 10^{-9}$ eV,  respectively.

\subsection{Displaced vertex search for $\Sigma^0$ at MATHUSLA}

The recently, proposed MATHUSLA \cite{Chou:2016lxi} is purposefully designed to detect  neutral long-lived particles produced at the LHC, 
and will be located $\sim 100$ m away from the beam collision point with an angle $\theta = [0.38, 0.8]$ from the LHC beam direction. 
Hence, it is ideally suited to detect long-lived particles with their decay lengths in the range of few hundred meters or even longer \cite{Curtin:2017izq}. 
If the particle decays with final state charged leptons and/or jets, 
the event can be easily reconstructed because the background SM events will be almost zero.

As shown in Fig.~\ref{fig:NH} (\ref{fig:IH}),  the decay length of ${\tilde \Sigma^0}$ corresponding to NH (IH), respectively, 
have decay length of 100 m for the benchmark values $\theta_{1,2,3} =0$,  $m_{{\tilde \Sigma}_1} = 500$ GeV,  $m_{\Sigma_{2,3}} (m_{\Sigma_{1,2}}) = 1$ TeV and $m_{1(3)}  = 10^{-9}$ eV.  
In the following, 
we investigate a possibility of detecting ${\tilde \Sigma}^0$ at MATHUSLA. 
At LHC, ${\tilde \Sigma}^0$ is produced in two ways. 
One is from ${\tilde \Sigma}^\pm$ production through a $s$-channel $Z/\gamma$ in the left diagram of Fig.~\ref{fig:prod},  
and the other is through  $s$-channel $W^\pm$ exchange process in the middle diagram of Fig.~\ref{fig:prod}.
In the first case, ${\tilde \Sigma}^0$ is produced by ${\tilde \Sigma}^\pm$ decay while in the other one ${\tilde \Sigma}^0$ is directly produced. 
The other particles produced from ${\tilde \Sigma}^\pm$ decay are soft and will not be considered in our  analysis. 
Since the MATHUSLA detector is very efficient in detecting long-lived particles, 
we will simply require that ${\tilde \Sigma}^0$ particle produced at the LHC reaches the MATHUSLA detector and decays inside the detector. 
In the following, 
we assume that the decaying ${\tilde \Sigma}^0$ particle can be detected with 100\% efficiency by MATHUSLA because the SM backgrounds are expected to be negligible.

Let us assume that ${\tilde \Sigma}^0$ produced at the LHC is emitted at an angle of $\theta$ away from the LHC beam line such that it decays after traveling a distance of $D_{\rm M}$.  
The decay length in the laboratory frame is given by 
\bea
c \tau_{\rm lab} = \frac {D_M}{\beta}.   
\eea 
Here, the $\beta$ is the boost factor for ${\tilde \Sigma}^0$ particle, 
\bea
\beta = \frac{1}{\sqrt {1 + \frac{{m_{{\tilde \Sigma}^0}}^2}{\overrightarrow{P}_{\rm tot}^2}}}, 
\eea
with a total momentum $\overrightarrow{P}_{\rm tot} = p_T \cosh {\eta}$, 
where $p_T$ is ${\tilde \Sigma}^0$'s transverse momentum and $\eta = - {\rm ln} \left[\tan(\theta/2)\right]$ is its rapidity. 
Hence, the proper decay length of ${{\tilde \Sigma}^0}$ is given by
\bea
c\tau_{{\tilde \Sigma}^0} = c \tau_{\rm lab}\times \sqrt{1 - \beta^2} = D_{\rm M} \frac {\sqrt{1 - \beta^2}}{\beta}. 
\eea
\begin{figure}[t!]
\begin{center}
\includegraphics[width=0.49\textwidth, height=5.5cm]{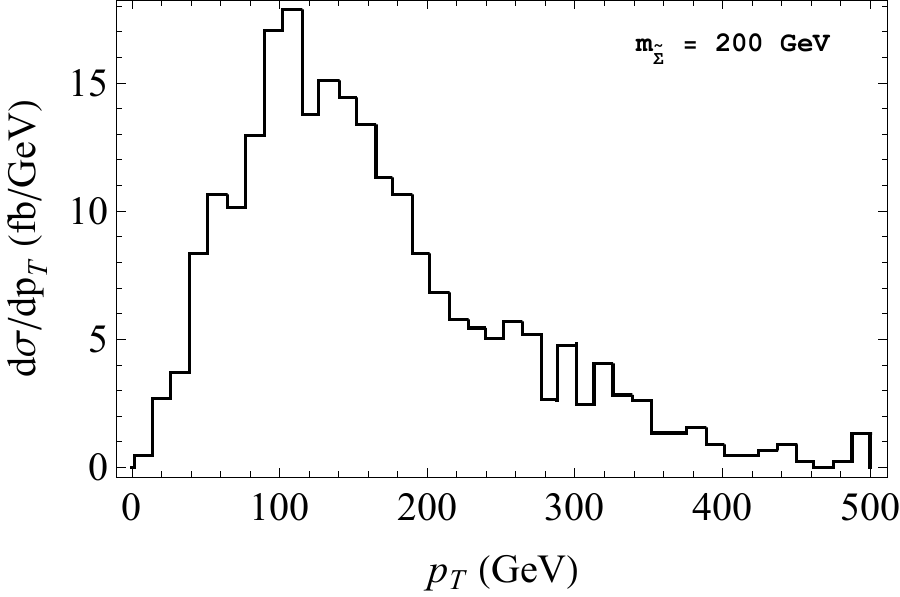}
\includegraphics[width=0.49\textwidth, height=5.5cm]{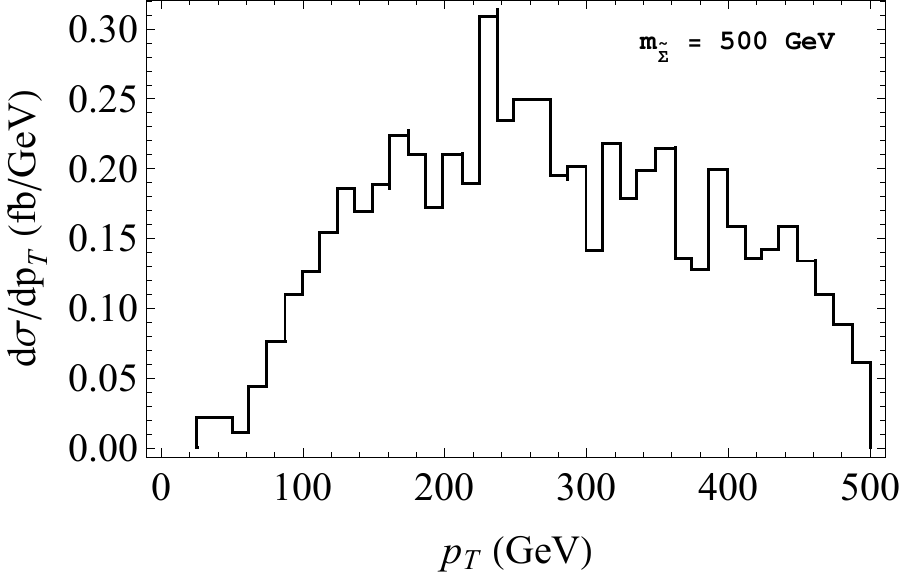}\\
\includegraphics[width=0.49\textwidth, height=5.9cm]{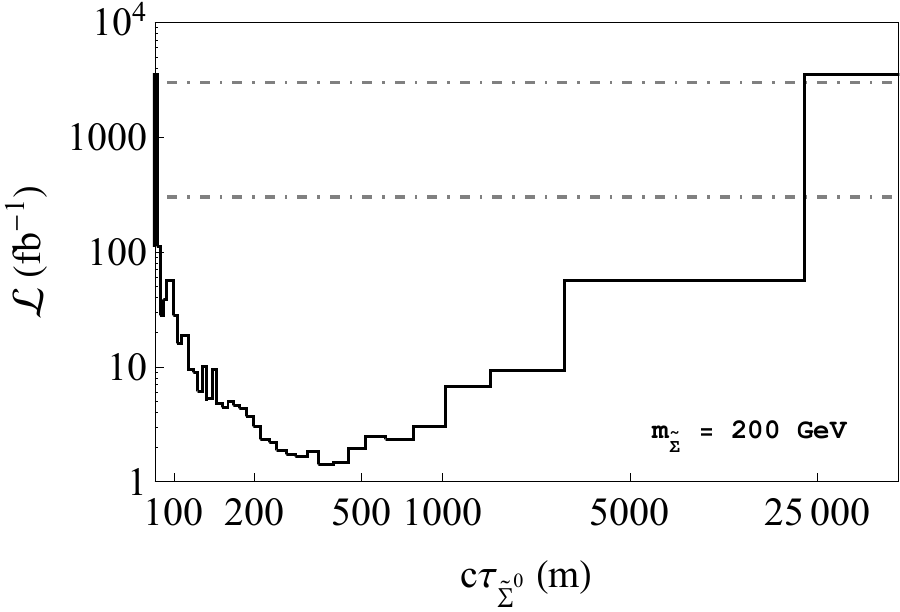}
\includegraphics[width=0.49\textwidth, height=5.9cm]{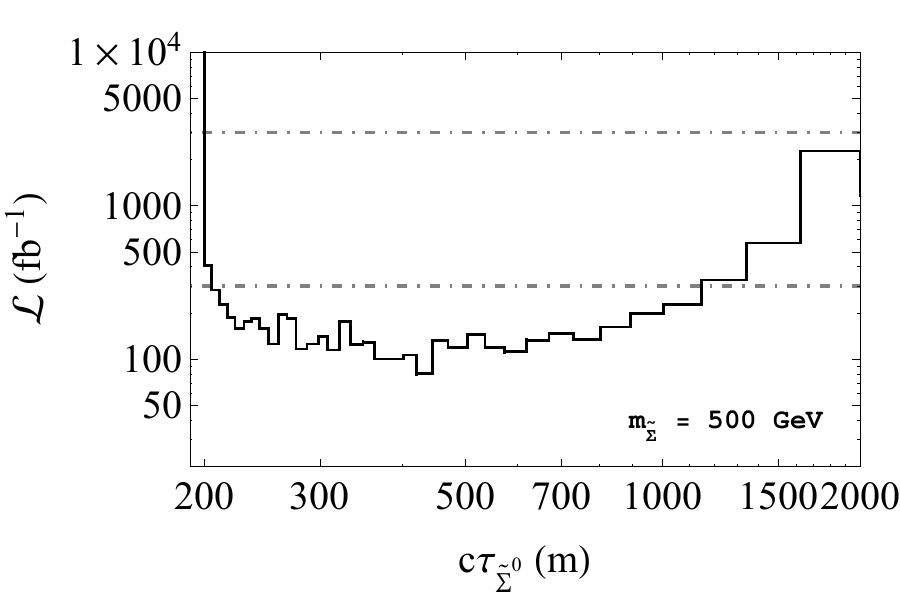}
\end{center}
 \caption{The top left (right) 
show the production cross section as a function of $p_T$ for $m_{\tilde \Sigma} = 200\;(500)$ GeV. 
In the bottom panels, we show the luminosities required to produce 25 ${\tilde \Sigma}$ at LHC which are directed towards MATHUSLA, and the horizontal top (bottom) line denote ${\cal L} = 300 \; (3000)$ fb$^{-1}$. }
\label{fig:mathspt}
\end{figure}
We fix $D_M = 100$ m and set $\theta \simeq 0.5$ corresponding to the location of the MATHUSLA detector. 
Hence, $P_{\rm tot} \simeq 2 p_T$ and 
\bea
c\tau_{{\tilde \Sigma}^0} \simeq 2D_{\rm M} \left(\frac{m_{{\tilde \Sigma}^0}}{p_T}\right) = 200 \left(\frac{m_{{\tilde \Sigma}^0}}{p_T}\right). 
\label{eq:DL}
\eea

To simulate the production of ${\tilde \Sigma}^0$ observed by MATHUSLA,
we employ {\sc MadGraph5aMC} {\sc @NLO}~\cite{Alwall:2011uj,Alwall:2014hca} and calculate the production cross section ($\sigma$) for the process. 
Since $c\tau_{{\tilde \Sigma}^\pm} \simeq 5$ cm for our benchmark is smaller than the benchmark value of 9 cm used in the Higgsino search analysis, we expect the lower-bound on ${\tilde \Sigma}^0$ mass to be weaker than $m_{{\tilde \Sigma}^0} \simeq 380$ GeV shown in the left panel of Fig.~\ref{fig:distrack}. 
Thus, in the following, we consider two benchmark values, $m_{{\tilde \Sigma}^0} = 200 \;(500)$ GeV. 
In the top left (right) panel of Fig.~\ref{fig:mathspt}, 
for $m_{{\tilde \Sigma}^0} = 200 \;(500)$ GeV, 
we show the cross section as a function of transverse momentum ($p_T$) of ${\tilde \Sigma}_0$. 
Because of low background, 
let us require 25 events ($N=25$) for a ``${\tilde \Sigma}^0$ discovery" at MATHUSLA. 
In the bottom left (right) panel of Fig.~\ref{fig:mathspt}, 
for $m_{{\tilde \Sigma}^0} = 200 \;(500)$ GeV, 
we show the luminosity (${\cal L} = N/\sigma$) required for the  discovery of ${{\tilde \Sigma}^0}$ as a function of its decay length, which we evaluated by using Eq.~(\ref{eq:DL}). 
The top (bottom) horizontal line denotes the integrated luminosity at HL-LHC with ${\cal L} = 300 \; (3000)$ fb$^{-1}$. 
From the bottom panels we find that the MATHUSLA can discover ${{\tilde \Sigma}^0}$ with a mass $200$ GeV and $500$ GeV, 
if their decay lengths are in the range ${\cal O} (10^2)$ m - ${\cal O}(10^4)$ m and ${\cal O} (10^2)$ m - ${\cal O}(10^3)$ m, respectively. 
Since the cross section decreases with increasing $m_{\tilde \Sigma}$ values, 
a heavier ${\tilde \Sigma}_0$ requires a higher luminosity. 
For example, 
we find that $m_{{\tilde \Sigma}^0} = 1000$ GeV requires ${\cal L} > 3000$ fb$^{-1}$ for any $c\tau$ values.

\subsection{Displaced vertex searches for $\Sigma^\pm$ at LHeC and FCC-he}

\begin{figure}[t!]
\begin{center}
\includegraphics[width=0.49\textwidth, height=5.5cm]{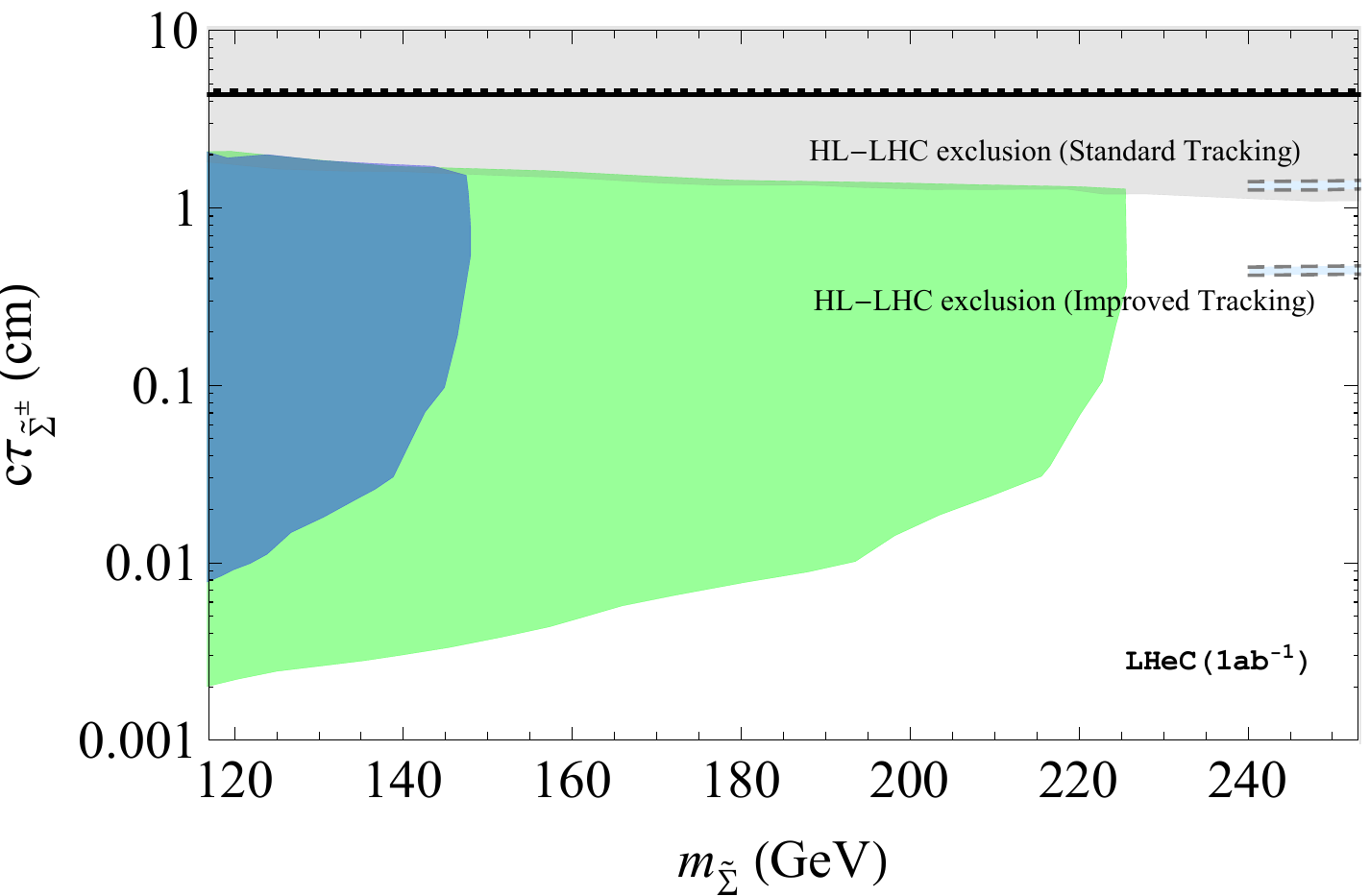}\\
\includegraphics[width=0.49\textwidth, height=5.5cm]{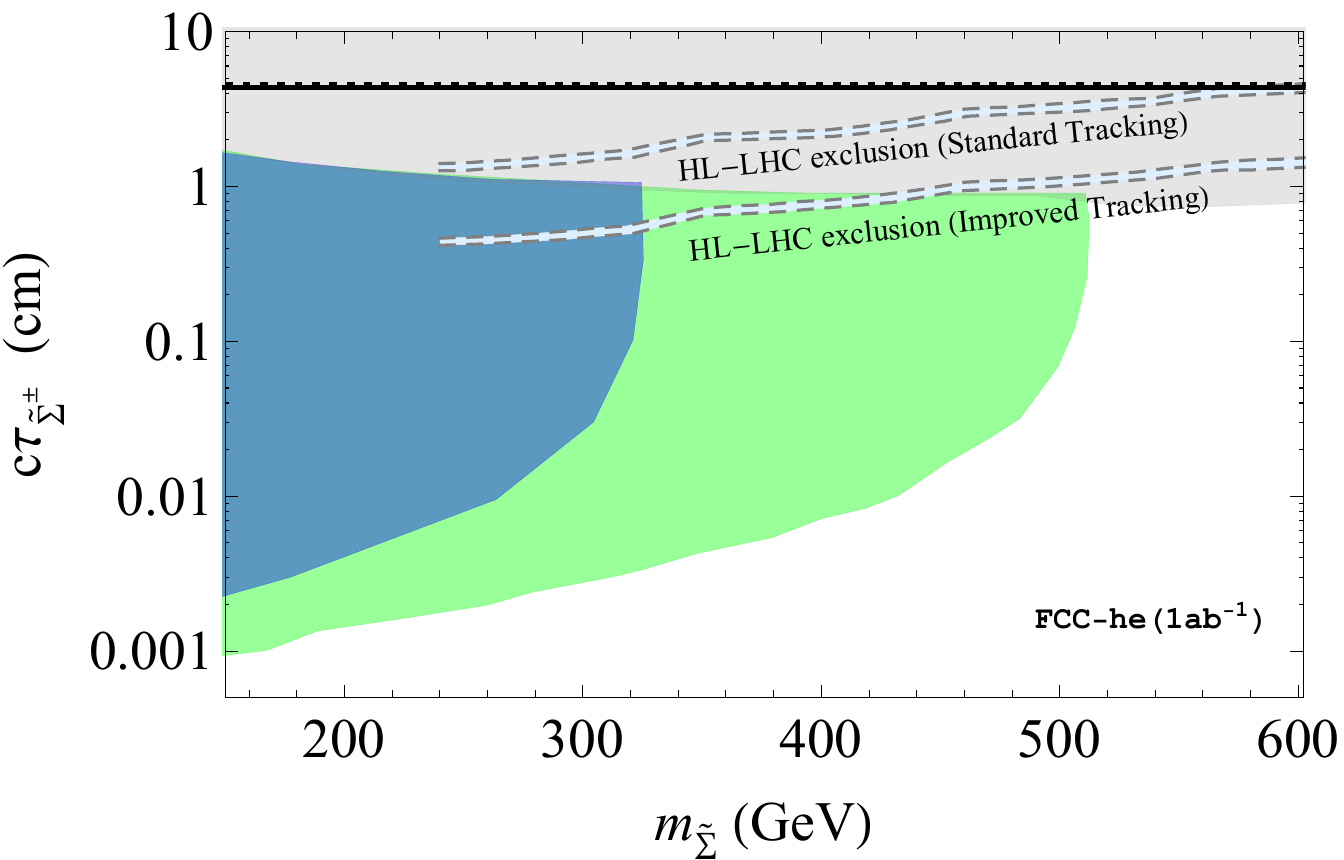}
\includegraphics[width=0.49\textwidth, height=5.5cm]{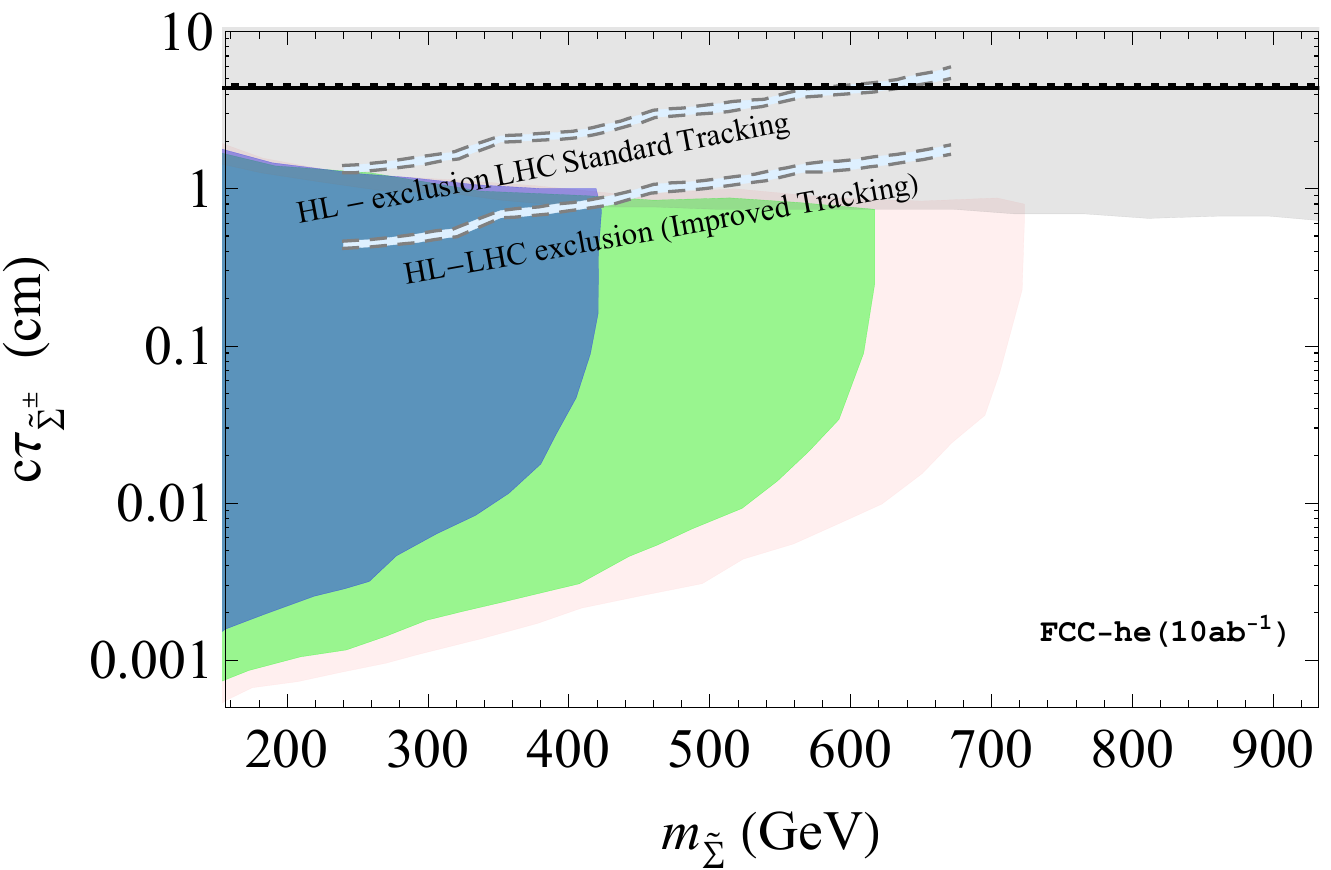}
\end{center}
 \caption{The top panel and two bottom panels show the prospects of observing displaced vertex from $\Sigma^\pm$ decay at LHeC and FCC-he, respectively. 
 The blue and green shaded regions indicate the parameter space where 10 (100) events with at least one long-lived $\Sigma^\pm$ observed at the LHeC/FCC-ec. 
 The light-red shaded region indicate 2-$\sigma$ exclusion sensitivity if the backgrounds are harder to reject. The grey shaded region are still allowed in our type-III seesaw scenario.
In all the panels, the solid (dotted) horizontal black lines are the decay lengths of ${\tilde \Sigma^\pm}$ corresponding to NH (IH), respectively, for the benchmark values $\theta_{1,2,3} =0$,  $m_{{\tilde \Sigma}_1} = 500$ GeV,  $m_{\Sigma_{2,3}} (m_{\Sigma_{1,2}}) = 1$ TeV and and $m_{1(3)}  = 10^{-9}$ eV.  }
\label{fig:epprospect}
\end{figure}

Let us now consider the displaced vertex signature arising from ${\tilde \Sigma}^{\pm}$ decay at $ep$-colliders\footnote{The prospect for probing the type-I seesaw scenario at LHeC has been investigated in Ref. \cite{das1} with prompt decays of the right-handed neutrinos.}. 
Authors in Ref.~\cite{Curtin:2017bxr} have studied the prospect to detect charged Higgsinos at LHeC and FCC-he experiments to show that these experiments can probe much shorter decay length, $10^{-3}$-$10$ cm, compared to the $pp$-colliders. 
As in the case of HL-LHC, we can recast their search prospect by exploiting the one-to-one correspondence between the Higgsino mass, its production rate and lifetime while taking into account the differences in the production rate of Higssino and ${\tilde \Sigma}$.  
Because the charged Higgsino decays 100\% to the long-lived neutral Higssino that evades detection and soft pion/SM leptons, 
the results in Ref.~\cite{Curtin:2017bxr} is only applicable to the type-III seesaw in the limiting case $m_{1 (3)} \to 0$ for NH (IH). 
See, for example, Figs.~\ref{fig:NH} and \ref{fig:IH}. 
Particularly, we fix $m_{1(3)}  = 10^{-9}$ eV for the following analysis.

Employing the prospect for detecting the long-lived charged Higgsino presented in Ref.~\cite{Curtin:2017bxr}, we show the prospects for LHeC (top panel) and FCC-he (bottom panels) in Fig.~\ref{fig:epprospect} along with the corresponding luminosities. 
In each plots, 
the blue and green shaded regions indicate the parameter space where 10 (100) events with at least one long-lived $\Sigma^\pm$ is observed at the LHeC/FCC-he; the green shaded regions correspond to 2-$\sigma$ exclusion sensitivity. 
In the bottom right plot, 
the red shaded regions 2-$\sigma$ exclusion sensitivity if the backgrounds are harder to reject.
The grey shaded region in all the plots is excluded for the long-lived Higgsino scenario but is still allowed in our scenario.  
In each panels of Fig.~\ref{fig:epprospect}, we also show the 5-$\sigma$ exclusion reach of HL-LHC obtained in the right panel of Fig.~\ref{fig:distrack} from the disappearing track searches. 
We also show the decay length of ${\tilde \Sigma^\pm}$ corresponding to NH (IH), which are depicted as the solid (dotted) horizontal black lines; the solid (dotted) lines are obtained for 
the benchmark values $\theta_{1,2,3} =0$,  $m_{{\tilde \Sigma}_1} = 500$ GeV and $m_{\Sigma_{2,3}} (m_{\Sigma_{1,2}}) = 1$ TeV and $m_{1(3)}  = 10^{-9}$ eV, respectively.

\section{Summary}\label{sec:con}
We have investigated the prospect of probing the type III seesaw neutrino mass generation mechanism at various collider experiments by searching for a disappearing track signature as well as a displaced vertex signature originating from the decay of $SU(2)_L$ triplet fermion $\Sigma$. 
We consider scenario with three triplet fermions, $\Sigma_{i=1,2,3}$, each of which include one neutral component, $\Sigma^0$, and two charged components, $\Sigma^\pm$. 
Because $\Sigma_i$ is primarily produced at colliders through their electroweak gauge interactions, 
its production rate is uniquely determined by its mass $m_{\Sigma_i}$.    
The type-III seesaw mechanism generates the observed neutrinos masses, which together with $m_{\Sigma_i}$, determine the decay length of $\Sigma^{0,\pm}_i$. 
We have shown that the neutral lepton associated with the lightest observed neutrino, particularly $\Sigma^0_{1 (3)}$ for NH (IH), respectively, can have lifetime $c\tau \gg 10$ m and evade detection at collider while their respective charged partners $\Sigma^\pm_{1 (3)}$ have a much shorter lifetime and decay inside the detector. 
The ${\Sigma}^\pm_{1 (3)}$ with a mass few hundered GeV and associated with the lightest observed neutrino $m_{1(3)}$ for NH (IH), respectively, decay and produce a disappearing track signature which can be searched at future $pp$-collider such as the HL-LHC.  
If the lightest observed neutrino, $m_{1(3)}$ for NH (IH), respectively, has mass around $10^{-9}$ eV, we have shown that both ${ \Sigma}^0_{1(3)}$ have a decay length of  around 100m and can be discovered at the proposed MATHUSLA detector. 
On the other hand, for NH (IH), we have also shown that ${\Sigma}^\pm_{1 (3)}$,  respectively, decay to produce an observable  displaced vertex signature at the LHeC and FCC-he.

\section{Acknowledgement}
We thank Oliver Fischer for helpful discussions. This work is supported in part by the United States Department of Energy grant de-sc0012447 (N.O) and de-sc0013880
(D.R).  

\end{document}